\documentclass[]{aa}

\usepackage{graphicx}

\usepackage{txfonts}

\usepackage{hyperref}
\usepackage{graphicx}
\usepackage{gensymb}
\usepackage{multirow}
\usepackage{amsmath}
\usepackage{tablefootnote}
\usepackage{lineno}
\usepackage{textcomp}
\usepackage{gensymb}
\linenumbers

\hypersetup{
    colorlinks=true,
    linkcolor=blue,
    citecolor=blue,
    urlcolor=blue,
    filecolor=blue
}

\begin{document} 

   \title{The HERON Coma Project: Exploring the low-surface-brightness frontiers in the Coma cluster}

   \titlerunning{The HERON Coma Project}
   \authorrunning{Román et al. }

   \author{Javier Román \inst{1} \corrauth{jromanastro@gmail.com}
   \and R.~Michael~Rich \inst{2} \email{rmrastro@gmail.com}
   \and Pablo~M.~Sánchez-Alarcón \email{pmsa.astro@gmail.com} \inst{3}
   \and Yolanda Jim\'enez-Teja \email{yojite@iaa.csic.es} \inst{4}
   \and Chester Li \email{zhuofu@uw.edu} \inst{2,5}
   \and Mónica Esquide \email{moniesquide@gmail.com} \inst{1}}

   \institute{Departamento de F\'isica, Universidad de Córdoba, Campus de Rabanales, Edificio Albert Einstein, E-14071 Córdoba, Spain
   \and Department of Physics \& Astronomy, University of California Los Angeles, 430 Portola Plaza, Los Angeles, CA 90095-1547, USA
   \and NASA Ames Research Center, Moffett Field, CA 94035, USA
   \and Instituto de Astrof\'isica de Andaluc\'ia--CSIC, Glorieta de la Astronom\'ia s/n, E--18008 Granada, Spain
   \and Department of Astronomy, University of Washington, 3910 15th Avenue, NE,  Seattle, WA, 98195, USA
   }
   \date{\today}

  \abstract
   {The technical complexity of data processing and residual instrumental systematics remain among the main barriers to reaching the faintest low-surface-brightness (LSB) regimes in the local Universe. Here, we present the HERON Coma Project, a deep $g$- and $r$-band imaging campaign on the Coma cluster carried out with the LSB-optimized 71-cm Jeanne Rich Telescope. The observing and processing strategy combines accurate flat-fielding, sky-background modeling and extended point spread function (PSF) characterization. As representative scientific applications, we explore the complex associated with the giant elliptical galaxies NGC~4839 and NGC~4816, in the southwestern region of Coma. Spatially coherent emission is recovered in direct imaging to $\mu_{\overline{g+r}}\simeq30~\mathrm{mag\,arcsec^{-2}}$, consistent with the nominal sensitivity inferred from the scaling of the background fluctuations. The PSF-deconvolved photometric profiles of NGC~4839 and NGC~4816 are traced to generalized isophotal radii of 705 and 329~kpc, reaching  $\mu_{\overline{g+r}}=32.2$ and $31.6~\mathrm{mag\,arcsec^{-2}}$, with data-derived signal-to-noise ratios of 4.1 and 7.0, respectively. The agreement between direct detection and the predictions from background fluctuations provides no clear evidence of a systematics-dominated detection floor over the surface brightness range and spatial scales examined. We also assess the observational challenges involved in extending reliable measurements to still fainter surface brightness levels. The HERON Coma Project opens a new window onto the faint stellar record of the complex and ongoing assembly of Coma, while demonstrating the potential of modest ground-based telescopes with suppressed instrumental reflections to provide highly competitive measurements of extended diffuse structures. We make the final calibrated images publicly available to the community.
   }


   \maketitle
\nolinenumbers

\section{Introduction} \label{sec:intro}

In the $\Lambda$ cold dark matter ($\Lambda$CDM) framework, dark matter halos grow hierarchically through accretion and merging, and galaxies inherit this assembly history \citep[e.g.,][]{Kauffmann1999, Cole2000, Bower2006, Johansson2012}. Galaxy properties consequently vary systematically with environment, including morphology, star formation activity, gas content, and stellar populations \citep[e.g.,][]{Dressler1980, Zabludoff1998, Balogh2004, Thomas2005, Baldry2006, Peng2010, Cooper2012, DominguezGomez2023}. Simulations reproduce many of these trends and emphasize the importance of preprocessing and satellite quenching \citep[e.g.,][]{Donnari2021}. Galaxy clusters therefore provide valuable laboratories for studying environmentally driven evolution through hydrodynamical, tidal, and preprocessing mechanisms \citep[e.g.,][]{Gunn1972, Larson1980, Moore1996, Balogh2000, Boselli2006}.

A substantial fraction of this stellar record lies in the low-surface-brightness (LSB) regime. Tidal debris, the extended envelopes of massive cluster galaxies, and the intracluster light (ICL) trace the stripping, merging, and redistribution of stars during cluster assembly \citep[e.g.,][]{Rudick2006, Cooper2015, Mihos2017, Kluge2021, Montes2022}. The least phase-mixed structures preserve signatures of recent interactions and, through their morphology, spatial distribution, colors, and stellar populations, constrain their progenitors and formation environments. Their observed properties provide sensitive tests of hierarchical assembly and galaxy accretion histories \citep[e.g.,][]{Cooper2010, Rey2022, Martin2022, MiroCarretero2025}, while the morphology and luminosity of the ICL have also been investigated as proxies for the underlying cluster mass distribution \citep[e.g.,][]{Montes2019, SampaioSantos2021}.

Despite their scientific value, extended sources at very low surface brightness remain among the most challenging targets in observational astronomy. Unlike compact sources, for which longer exposures generally improve the achievable depth predictably, diffuse emission may become limited by systematic uncertainties rather than photon statistics \citep[see the review by][]{Mihos2019}. Flat-fielding residuals, sky-subtraction biases, scattered light from the extended point spread function (PSF), instrumental reflections, and unresolved background sources, among other matters, can all compromise LSB measurements. Diffuse Galactic cirrus introduces an additional astrophysical foreground whose optical emission can mimic the filamentary morphology of tidal debris and bias photometry of extended sources \citep[e.g.,][]{MivilleDeschenes2016, Roman2020, Smirnov2023, 2025A&A...704A.269L}.

Addressing these limitations has required improvements in source detection and masking \citep[e.g.,][]{Akhlaghi2015, Haigh2021}, high-precision sky background procedures \citep[e.g.,][]{Watkins2024}, and characterization of the extended PSF and instrumental scattered light \citep[e.g.,][]{Slater2009, Sandin2014, Karabal2017, InfanteSainz2020, GarateNunez2024, Sedighi2025}. Multi-wavelength information and dedicated decomposition techniques have likewise been developed to identify and mitigate Galactic cirrus contamination \citep{2023ApJ...953....7L,2025ApJ...979..175L}.

Observational efforts to probe this regime have followed two broad, partly overlapping approaches. Small-aperture, wide-field systems optimized for LSB imaging facilitate control of instrumental reflections and other systematics, allowing observations to approach the background-noise limit \citep[e.g.,][]{MartinezDelgado2010, Mihos2017, Abraham2014}. Large-aperture facilities instead exploit their greater collecting area together with advanced reduction techniques, although residual systematics may ultimately limit their performance \citep[e.g.,][]{Duc2015, 2015A&A...581A..10C, TrujilloFliri2016, 2018MNRAS.475.3348H, 2018RNAAS...2..144R, 2020ApJS..247...43K, Trujillo2021}. The Euclid Early Release Observations have also demonstrated the efficiency of space-based imaging when instrumental stability is combined with processing designed to preserve diffuse emission \citep{Cuillandre2025}.

The Coma cluster is a particularly important target for LSB studies. As one of the nearest massive clusters, it has played a central historical role in investigations of both the large-scale mass distribution and diffuse stellar emission. Its dynamics provided the first evidence of substantial unseen matter in galaxy clusters \citep{Zwicky1933}, and one of the earliest detections of ICL \citep{Zwicky1951}. Modern measurements confirm Coma as a very massive cluster, with an estimated mass of $M_{200c}=10^{15.10\pm0.15}\,h^{-1}\,M_{\odot}$ \citep{Ho2022}.

Coma is neither dynamically simple nor fully relaxed. Its central giant elliptical galaxies, NGC~4874 and NGC~4889, are associated with distinct dynamical components and a recent or ongoing merger in the cluster core \citep[e.g.,][]{Briel1992, Watt1992}. Additional substructures and infalling systems are present throughout the cluster, most notably the NGC~4839 group to the southwest \citep[e.g.,][]{Adami2005, Gerhard2007}. The galaxy distribution, hot intracluster medium, and diffuse stellar component consistently indicate continuing accretion and a complex assembly history \citep[e.g.,][]{JimenezTeja2019, Gu2020, Oh2023}, whose record is preserved in the LSB structures of Coma.

Coma also presents a particularly favorable foreground environment for deep optical imaging. The cluster lies close to the north Galactic pole, at a Galactic latitude of $b\simeq88^{\circ}$, where foreground extinction is very low and contamination by diffuse Galactic cirrus is expected to be comparatively faint. The low density of foreground Milky Way stars also reduces crowding. Nevertheless, Coma remains technically demanding because of its large angular extent and the presence of several bright stars close to the core region. The extended wings of their point-spread functions overlap with parts of the cluster emission and can hinder the detection of extremely faint structures, while the large fraction of the field occupied by astrophysical emission complicates the identification of genuinely source-free regions for sky-background modeling.

The Halos and Environments of Nearby Galaxies (HERON) survey is a deep-imaging program conducted with the LSB-optimized 71-cm Jeanne Rich Telescope to investigate faint structures in and around nearby galaxies \citep{Rich2017,Rich2019,Mosenkov2020}. Within this broader framework, the HERON Coma Project was designed to obtain a deep and spatially extended view of the LSB structures in the central and southwestern regions of the Coma cluster. The Jeanne Rich Telescope has also been used for studies of faint tidal structures and diffuse galaxies \citep[e.g.,][]{Muller2019,Gannon2021,Ogle2024}.

Two previous studies have already presented specific scientific results based on the HERON Coma Project. \citet{Roman2023} reported the discovery of the Giant Coma Stream, an extremely faint ($\mu_{g,max}$~$=$~29.5~mag~arcsec$^{-2}$) and narrow stellar structure extending for approximately 500~kpc and interpreted as the debris of a disrupted dwarf galaxy. \citet{JimenezTeja2025} subsequently presented a dedicated analysis of the Coma ICL, combining the HERON observations with complementary spectroscopic data to characterize its morphology and fractional contribution to the total cluster light, and to relate its diffuse structures to spectroscopically identified galaxy groups and the ongoing assembly of the cluster.

In this work, we present the complete HERON Coma observational campaign and describe the reduction and coaddition procedures developed to preserve extended emission at extremely low surface brightness. We characterize the dataset and present an overview of the diffuse structures detected across the observed field, including deep photometric profiles of NGC~4839 and NGC~4816. The final calibrated coadded images are made publicly available, as detailed in the Data availability section.

Throughout this work, we adopt a luminosity distance of $100\,\mathrm{Mpc}$ to the Coma cluster, corresponding to a distance modulus of $m-M=35.0$~mag \citep{2001ApJ...557L..31L}. At the cluster redshift, the cosmological parameters of \citet{Planck2020} give a physical scale of 0.462~kpc~arcsec$^{-1}$. The foreground Galactic extinction is very low, with $A_g=0.027$~mag and $A_r=0.019$~mag according to \citet{Schlafly2011}. Given these small values, no extinction correction is applied to the photometry. All magnitudes are given in the AB system.

\section{Observations}
\label{sec:observations}

The Jeanne Rich Telescope is located at the Polaris Observatory Association site in Lockwood Valley, near Frazier Park, California, at an elevation of approximately 1615~m. It is a 71-cm, $f/3.2$ prime-focus reflector equipped with a Ross doublet field corrector and a conical baffle with ring stops designed to suppress scattered light. This simple optical configuration is particularly well suited to imaging extended LSB emission. For the HERON Coma campaign, the telescope was equipped with a Finger Lakes Instruments ML09000 CCD with 12~$\mu$m pixels. The resulting pixel scale is 1.114~arcsec~pixel$^{-1}$, providing a field of view of approximately $56.6\times56.6$~arcmin$^{2}$ \citep[see][]{Brosch2015,Rich2019}. The observations presented here were obtained through Sloan $g$ and $r$ filters, with nominal effective wavelengths of approximately $477$ and $623\,\mathrm{nm}$ and approximate wavelength ranges of $409$--$546\,\mathrm{nm}$ and $555$--$692\,\mathrm{nm}$, respectively.

The observing strategy for the HERON Coma Project was designed to facilitate the construction of science superflats. A wide dithering pattern was adopted so that astronomical sources were projected onto different regions of the detector throughout the observing sequence. After masking the detected sources and normalizing the individual exposures, the large number of dithered frames could therefore be combined to construct high-signal-to-noise flat-field images. The construction and application of these superflats are described in Sect.~\ref{sec:Data}.

The footprint was designed primarily to map the diffuse stellar emission in the central region of Coma, which is dominated by the giant elliptical galaxies NGC~4874 and NGC~4889 and the surrounding ICL. The observations also cover the massive early-type galaxies NGC~4839 and NGC~4816, located to the southwest, together with the broader environment surrounding the cluster core. Given the approximately 56.6-arcmin field of view of the instrument, we constructed a mosaic covering roughly $1.5^{\circ}\times1.5^{\circ}$, corresponding to projected dimensions of approximately $2.5\times2.5$~Mpc at the adopted distance of Coma. The mosaic is centered at right ascension (R.A.) $=194.75^{\circ}$ and declination (Dec.) $=+27.85^{\circ}$. The overlapping pointing pattern was designed to maximize the integration time in the central region, with the exposure depth decreasing progressively toward the boundaries of the footprint.

All individual images were obtained with an exposure time of 300~s in both filters. This duration provided a compromise between maximizing the signal accumulated in each exposure and maintaining reliable telescope guiding. The use of two photometric bands provides color information and a useful consistency check on the faintest detected structures. In particular, morphological agreement between features independently recovered in the $g$- and $r$-band mosaics reduces the likelihood that they are produced by filter-dependent reflections or reduction artifacts. After photometric calibration to a common zero point, the two bands were combined through their arithmetic mean to construct a higher-signal-to-noise $\overline{g+r}$ detection image.

The observations were scheduled preferentially during dark or low-background conditions. The presence of the Moon was avoided during the $g$-band campaign, whereas some $r$-band exposures were obtained under low lunar illumination. Atmospheric seeing was not a primary selection criterion because the structures of interest, including stellar halos and diffuse features, have angular scales substantially larger than the PSF. Nevertheless, observations under particularly poor image quality, with seeing worse than approximately 5~arcsec, were generally avoided. Priority was otherwise given to maximizing the accumulated integration time under suitably dark and stable conditions.

The $g$-band observations were obtained between March and June 2019, while the $r$-band campaign was carried out between January and June 2020. In total, the campaign acquired 566 exposures of 300~s in the $g$ band, corresponding to 47.2~h of accumulated exposure time, and 635 exposures of 300~s in the $r$ band, corresponding to 52.9~h. These values include all acquired exposures before the quality-control selection described in Sect.~\ref{sec:Data} and do not represent the integration time reached at any individual position within the mosaic.

\section{Data reduction and image characterization}
\label{sec:Data}

\subsection{Basic reduction and flat-field construction}
\label{sec:reduction}

The initial CCD processing consisted of bias and dark-current subtraction, followed by flat-field correction. Bias and dark exposures were obtained at both the beginning and the end of each observing night, with typically about 40 bias frames and 15 dark frames acquired in total per night. Master bias and dark frames were constructed by combining the corresponding exposures using a $3\sigma$-clipped mean and were subsequently subtracted from the science images.

Science superflats were constructed directly from the bias- and dark-subtracted exposures. Astronomical sources were aggressively masked by combining detections from \texttt{SExtractor} \citep{Bertin1996} and \texttt{NoiseChisel} \citep{Akhlaghi2015}. \texttt{SExtractor} was run with \texttt{DETECT\_THRESH}=0.7, corresponding to a detection threshold of $0.7\sigma$ above the local background, and was used primarily to identify compact sources. \texttt{NoiseChisel} was run with \texttt{qthresh}=0.3, which defines the quantile threshold applied to the convolved image, and provided greater sensitivity to diffuse and extended LSB emission. The masked images were normalized to a common flux and combined using a $3\sigma$-clipped mean to produce a first-pass superflat. This preliminary flat was applied to the science exposures solely to improve the detection and masking of faint and extended sources. The improved masks obtained from these corrected images were then transferred back to the original bias- and dark-subtracted exposures, which were again normalized and combined. The resulting second-iteration superflat was adopted as the final flat-field correction.

To minimize temporal variations in the illumination pattern and maximize the accuracy of the flat-field correction, the observations were divided into sets hereafter referred to as \emph{runs}. Each run comprised all observations acquired between successive bright-Moon periods, typically spanning 10--20 nights depending on weather conditions and target visibility. During each run, the corresponding filter remained installed and fixed at the same position in the filter wheel. This strategy reduced the possibility of introducing flat-field variations associated with changes in the optical configuration.

The superflat for each run and filter was constructed using all suitable exposures obtained with the telescope during the corresponding period, including observations acquired for programs other than the HERON Coma Project. Because these images sampled different regions of the sky, their inclusion increased the signal-to-noise ratio of the superflat and reduced the residual imprint of similar astronomical sources. A separate final superflat was therefore generated for each observing run and photometric band and applied to all HERON Coma exposures acquired within that run and filter.

\subsection{Astrometric calibration, background modeling, and coaddition}
\label{sec:coaddition}

Accurate background subtraction and image coaddition are essential for preserving extended emission in the LSB regime. Even after the flat-field correction described above, individual exposures contain spatially varying backgrounds produced by airglow, scattered light, light pollution, and residual instrumental signatures. We therefore developed an iterative procedure in which source masking, background modeling, and coaddition were progressively refined.

An initial astrometric solution was derived for each exposure using \texttt{Astrometry.net} \citep{Lang2010} and subsequently refined with \texttt{SCAMP} \citep{Bertin2006} to obtain the final high-precision astrometric calibration. The individual images were then resampled onto a common astrometric grid using \texttt{SWarp} \citep{Bertin2002} with the \texttt{LANCZOS3} interpolation kernel, chosen to minimize degradation of the spatial resolution during reprojection. To construct the initial, or seed, coadd, sources in each exposure were aggressively masked by combining detections from \texttt{SExtractor} and \texttt{NoiseChisel}. A single constant sky level, calculated as a $3\sigma$-clipped mean of the unmasked pixels in each exposure, was subtracted from each masked exposure before coaddition. Restricting the first-pass background subtraction to a constant minimized the risk of removing genuine large-scale emission, although residual gradients consequently remained in the seed mosaic.

We iteratively created mosaics improving the steps associated with the background calibration. At each iteration, a new source mask was generated from the coadd produced in the preceding step. Because the effective depth, and hence the pixel noise, varies across the footprint, the detection image was constructed by multiplying the coadd by the square root of its weight map. When the weight map is proportional to the inverse variance, this produces an approximately signal-to-noise-weighted image and allows a more uniform detection threshold to be applied across regions with different exposure depths, avoiding artifacts in the detection maps. Before running \texttt{NoiseChisel}, this detection image was spatially binned by $2\times2$ pixels. The binning increases the signal-to-noise ratio of extended low-surface-brightness emission and therefore improves the sensitivity of \texttt{NoiseChisel} to diffuse structures that remain difficult to detect at the native pixel scale. The resulting segmentation mask was subsequently expanded back to the native $1\times1$ sampling and registered on the original coadd grid. The coadd-based mask therefore identifies sources and diffuse emission at substantially fainter levels than masks derived from individual exposures. The coadd and its associated mask were iteratively refined until no further substantial changes were observed.

Within each iteration, the individual exposures were photometrically calibrated against the Dark Energy Camera Legacy Survey \citep[DECaLS;][]{Dey2019} in the $g$ and $r$ bands and scaled to a common photometric zero point of ZP~=~22.5~mag. After reprojection onto the common astrometric grid, the modeled halos of bright stars were subtracted from each exposure. The construction of the extended PSF model and the model of stars in the field are described in Appendices \ref{sec:PSF} and \ref{sec:starsubtraction}, respectively. Removing the extended scattered-light contribution from bright stars before estimating the sky background reduces the area contaminated by stellar halos, leaving a larger fraction of the image available for the subsequent sky-background modeling, while preventing this light from being partially absorbed into the fitted background model \citep{Watkins2024}.

The residual sky background in each masked exposure was represented using a two-dimensional Zernike polynomial \citep{Zernike1934, Lakshminarayanan2011}. Zernike polynomials provide a compact basis in which the complexity of the fitted background can be controlled through the adopted maximum radial order, $n_{\mathrm{max}}$. Originally developed to describe wavefront aberrations over a circular optical pupil, their lowest-order modes naturally represent smooth, large-scale spatial patterns. The $n=0$ mode corresponds to a constant piston term, whereas the $n=1$ mode describes linear tip and tilt; higher orders progressively introduce additional low-spatial-frequency structure. This makes low-order Zernike expansions well suited to modeling the smooth gradients produced by illumination variations, while reducing their ability to reproduce the irregular and spatially complex structures expected from genuine diffuse emission associated with galaxies and the cluster. Zernike polynomials have also been used in recent astronomical background-modeling applications \citep{Zhang2023, SanchezAlarcon2023, Perez2024}. 

We initially attempted to model the background of every exposure using a low-order expansion with $n_{\mathrm{max}}=1$. When residual large-scale gradients remained, the maximum radial order was increased as required, initially up to $n_{\mathrm{max}}=3$. Although the iterative procedure substantially improved both the source masks and the coadds, a subset of exposures produced persistent residual structures. These cases generally occurred when the source mask covered an entire detector quadrant or an even larger fraction of the image. Under these conditions, the background model was poorly constrained over the masked region and could extrapolate to unrealistic values, producing artificial gradients in the resulting coadd.

We therefore performed a final visually supervised pass using the deepest converged source mask. For each exposure, background models constructed with different maximum radial orders were inspected, and the lowest-order model that adequately removed the large-scale gradient without introducing conspicuous structured residuals was adopted. Orders of $n_{\mathrm{max}}\leq3$ were sufficient for the great majority of the exposures, while $n_{\mathrm{max}}=4$ was required in a small number of cases. A consequence of the deliberately restricted flexibility of the Zernike basis is that localized or irregular instrumental features that are not well represented by low-order modes remain visible after background subtraction. This was particularly evident for ghost reflections produced by bright stars outside the field of view. These features were therefore identified and masked manually whenever possible. However, because some reflections simply pass unnoticed during the procedure, a complete removal could not be guaranteed, and faint ghost-light features are expected to remain in the final mosaics. Exposures affected by gradients that were too strong or too complex to be modeled reliably were excluded from the final coaddition.

After subtraction of the adopted background model, the noise of each exposure was estimated by fitting a Gaussian function to the distribution of its unmasked pixel values. The fitted width, $\sigma_{\mathrm{sky}}$, was adopted as an empirical estimate of the background noise, and the relative statistical weight of each exposure was defined as proportional to $\sigma_{\mathrm{sky}}^{-2}$. The accepted sky-subtracted exposures were then combined by applying a $3\sigma$ rejection at each output pixel and calculating a weighted mean of the retained values.

In the $g$ band, 108 of the 566 acquired exposures were rejected, corresponding to 19\% of the dataset. The final $g$-band coadd therefore contains 458 exposures, with a summed accepted exposure time of 38.2~h. In the $r$ band, 175 of the 635 exposures were rejected, corresponding to 28\%, leaving 460 exposures and a summed accepted exposure time of 38.3~h. These values represent the total duration of all exposures included in each mosaic rather than the integration time reached at any individual position.

Although this quality selection reduces the nominal photon-limited depth relative to a stack containing every acquired frame, extended LSB measurements are particularly sensitive to residual large-scale background structure. Rejecting exposures with poorly constrained backgrounds therefore improves the fidelity of the final mosaics despite the reduction in accumulated exposure time. The mean PSF full width at half maximum (FWHM), estimated from the empirical PSF models described in Appendix~\ref{sec:PSF}, is 3.0~arcsec in the $g$ band and 3.6~arcsec in the $r$ band. The local FWHM is nevertheless expected to vary smoothly across the footprint because the relative contribution of individual exposures with different image quality changes with position.

\subsection{Analysis footprint and nominal depth}

The integration time and nominal limiting surface brightness vary across the HERON Coma footprint owing to the overlapping pointing pattern and the rejection of individual exposures during the quality-control procedure. The central region has the deepest and most homogeneous coverage, while the local integration time decreases progressively toward the boundaries of the mosaic. Smaller-scale variations differ between the two filters; in particular, the eastern part of the $r$-band mosaic has somewhat lower coverage than the corresponding region in the $g$ band. The corresponding integration-time and nominal-depth maps are presented in Appendix~\ref{sec:depth_appendix}.

The nominal limiting surface brightness was derived from the background noise measured at the native pixel scale and extrapolated to the equivalent surface brightness corresponding to a $3\sigma$ fluctuation in a $10\times10~\mathrm{arcsec}^{2}$ aperture, following \citet{Roman2020}. It therefore characterizes the random-noise sensitivity of the mosaics on this reference angular scale and should be distinguished from the fainter levels that can be reached in direct detection through spatial binning or azimuthal averaging. The complete procedure and the empirical relation between local integration time and nominal limiting surface brightness are described in Appendix~\ref{sec:depth_appendix}.

For the scientific analysis, we define a common footprint requiring a minimum local integration time of 1.5~h in both the $g$ and $r$ bands. This threshold corresponds to nominal limiting surface brightnesses of approximately $\mu_{\mathrm{lim},g}=28.3~\mathrm{mag\,arcsec^{-2}}$ and $\mu_{\mathrm{lim},r}=27.9~\mathrm{mag\,arcsec^{-2}}$, adopting the $3\sigma$ criterion in $10\times10~\mathrm{arcsec}^{2}$ apertures.

The maximum local integration times are 35.6~h in the $g$ band and 36.0~h in the $r$ band. These correspond to maximum nominal depths of $\mu_{\mathrm{lim},g}=30.1~\mathrm{mag\,arcsec^{-2}}$ and $\mu_{\mathrm{lim},r}=29.6~\mathrm{mag\,arcsec^{-2}}$, respectively, expressed in the equivalent $3\sigma$, $10\times10~\mathrm{arcsec}^{2}$ metric defined above. A substantial fraction of the central footprint reaches nominal depths of approximately $29.0~\mathrm{mag\,arcsec^{-2}}$ or fainter in both filters.

\section{Results}
\label{sec:results}

\subsection{Global low-surface-brightness view of Coma}
\label{sec:global_view}

\begin{figure*}
    \centering
    \includegraphics[width=\textwidth]{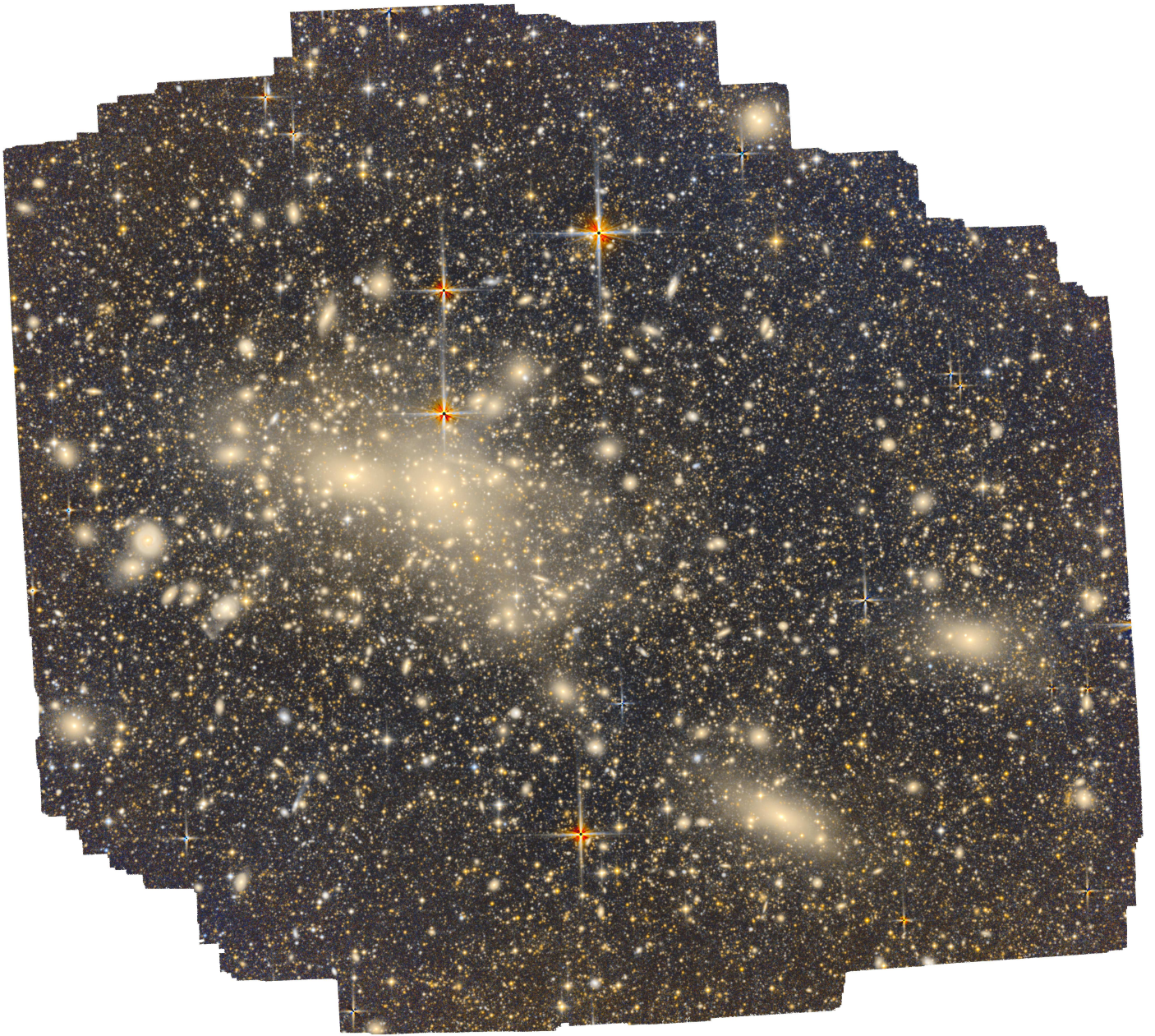}
    \caption{Color-composite image of the HERON Coma footprint constructed from the final $g$- and $r$-band mosaics at the native pixel scale of 1.114~arcsec~pixel$^{-1}$. North is up and east is to the left. The central region is dominated by the giant elliptical galaxies NGC~4889 and NGC~4874 and their surrounding diffuse ICL. The massive early-type galaxies NGC~4839 and NGC~4816 are visible in the southwestern part of the footprint.}
    \label{fig:Coma_color}
\end{figure*}

\begin{figure*}
    \centering
    \includegraphics[width=\textwidth]{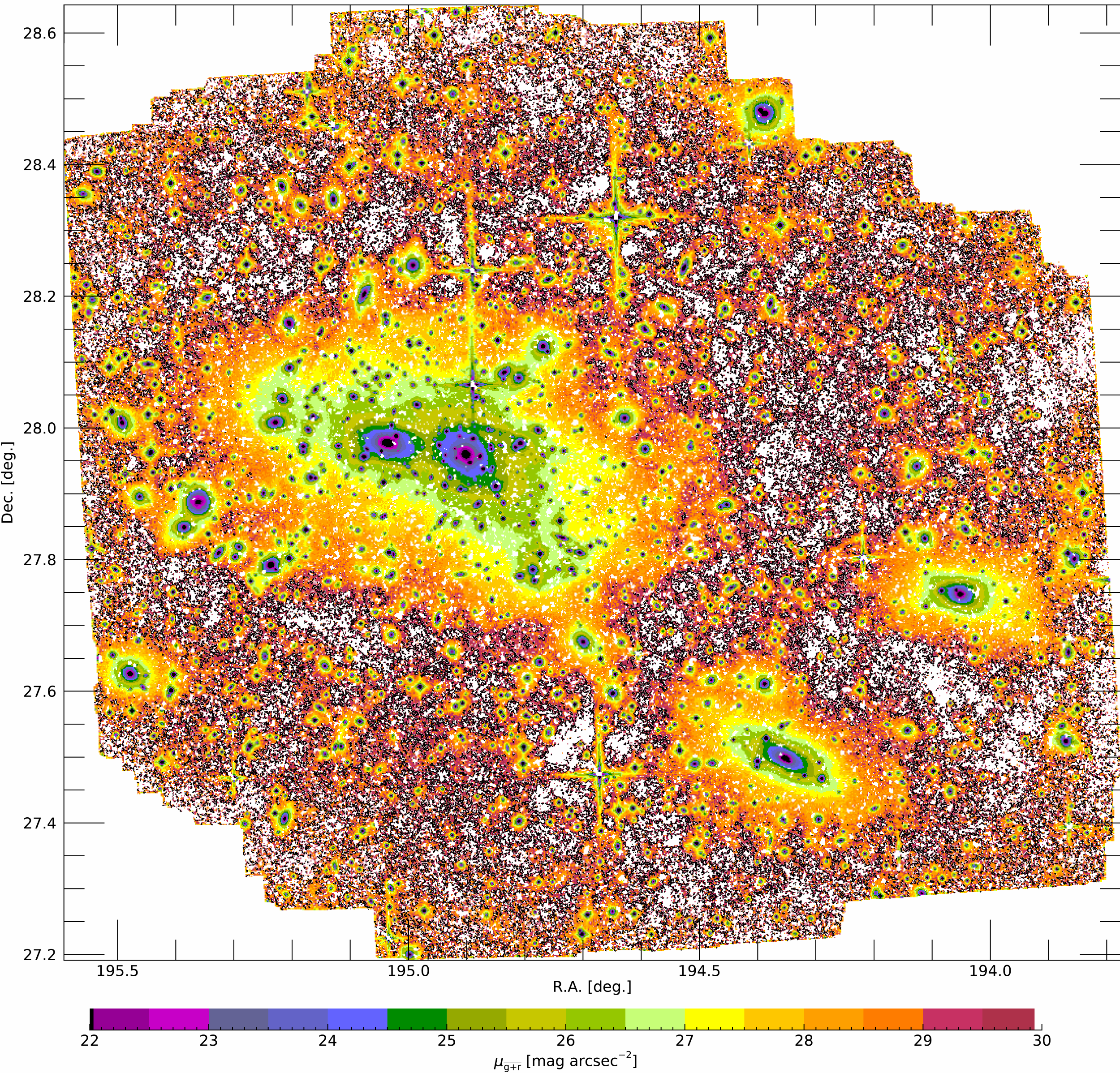}
    \caption{Enhanced surface-brightness image of the Coma cluster constructed from the arithmetic mean of the $g$- and $r$-band mosaics. Compact sources were masked, and the image was spatially binned to a pixel scale of 5.57~arcsec~pixel$^{-1}$. The color scale indicates surface brightness in units of mag~arcsec$^{-2}$. A black contour at $\mu_{\overline{g+r}}=30~\mathrm{mag\,arcsec^{-2}}$ is overlaid to facilitate visualization and delineate the faint diffuse structures.}
    \label{fig:Coma_mag}
\end{figure*}

Figure~\ref{fig:Coma_color} presents the final HERON view of the common analysis footprint, encompassing the central NGC~4874--NGC~4889 region inside the ICL of Coma and the NGC~4839--NGC~4816 complex to the southwest. To enhance the faintest diffuse structures, we constructed the surface-brightness map shown in Fig.~\ref{fig:Coma_mag} from the mean of the photometrically calibrated $g$- and $r$-band mosaics. Compact sources were identified with \texttt{SExtractor}, and detections with segmentation areas smaller than 950 native pixels were masked, removing most background galaxies and faint foreground stars while preserving extended emission. The masked $\overline{g+r}$ image was spatially binned by $5\times5$, yielding a final pixel scale of 5.57~arcsec~pixel$^{-1}$. Each output pixel was calculated using only unmasked input pixels. This processing enhances diffuse low-signal-to-noise emission at the expense of spatial resolution and reduces contamination from compact sources.

The adopted binning represents a compromise between surface-brightness sensitivity and the preservation of relatively small-scale structures. Rescaling the nominal depth relations derived in Appendix~\ref{sec:depth_appendix} to an individual $5.57\times5.57~\mathrm{arcsec}^{2}$ binned pixel gives nominal $3\sigma$ limits of $\mu_{\mathrm{lim},g}=29.4~\mathrm{mag\,arcsec^{-2}}$ and $\mu_{\mathrm{lim},r}=29.0~\mathrm{mag\,arcsec^{-2}}$ in the regions of maximum exposure time. Because the displayed image is constructed as the arithmetic mean of the $g$- and $r$-band mosaics, their independent random-noise contributions combine according to

\begin{equation}
\sigma_{\overline{g+r}}
 = \frac{1}{2}\sqrt{\sigma_g^2+\sigma_r^2},
\end{equation}

\noindent where $\sigma_g$ and $\sigma_r$ are the corresponding $1\sigma$ background dispersions at the binned-pixel scale. Propagating the band-specific fluctuations yields an effective nominal $3\sigma$ limit of $\mu_{\mathrm{lim},\overline{g+r}}=29.54~ \mathrm{mag\,arcsec^{-2}}$, approximately $0.12$~mag deeper than the $g$-band image alone, while reducing the color-dependent selection effects associated with detection in a single band. Consistently, in the deepest parts of Fig.~\ref{fig:Coma_mag}, spatially coherent diffuse emission can be visually traced to $\mu_{\overline{g+r}}\simeq29.5~\mathrm{mag\,arcsec^{-2}}$. The agreement between these values indicates that the observed detection level is broadly consistent with the nominal random-noise sensitivity, although localized residual systematics may remain present. Since the optimal binning depends on the characteristic angular scale of the emission under study, we adopt different binning factors for the analyses presented below to maximize the detectability of each feature while preserving the spatial information needed to characterize its morphology.

The enhanced image reveals diffuse stellar structures over a broad range of spatial scales. Within the central ICL region, at comparatively bright levels, the field contains numerous galaxy-scale halos, asymmetric envelopes, and relatively sharp tidal features. At fainter levels, these structures become less distinct and increasingly overlap with a smoother ICL component. This transition becomes particularly apparent around $\mu_{\overline{g+r}}\sim27.5~\mathrm{mag\,arcsec^{-2}}$, comparable to the surface-brightness thresholds of the BCG-ICL transition commonly adopted in observational studies \citep[e.g.,][]{Zibetti2005, Kluge2021, Montes2022}, also qualitatively consistent with simulations in which dynamically young ICL contains filaments and streams that gradually disperse \citep{Rudick2006,Cooper2015}. The southwestern footprint contains the extended stellar components of NGC~4839 and NGC~4816 and the diffuse structures examined in Sect.~\ref{sec:4839_4816}.

Some localized ghost-like artifacts are visible in the northern part of the footprint. Their instrumental origin is indicated by their presence in only one band, or by their substantially greater contrast in one band than in the other. A ghost residual is visible adjacent to the star at $(\mathrm{R.A.},\mathrm{Dec.})=(194.357^{\circ},+28.307^{\circ})$, while two additional diffuse residual patches occur near $(\mathrm{R.A.},\mathrm{Dec.})=(194.9^{\circ},+28.4^{\circ})$ and $(\mathrm{R.A.},\mathrm{Dec.})=(194.9^{\circ},+28.5^{\circ})$.

On larger spatial scales, sky oversubtraction would manifest as broad negative depressions around extended sources or as discontinuities across the mosaic. We find no clear evidence of such signatures around the stellar halos, the central ICL, or the diffuse emission surrounding NGC~4839 and NGC~4816, nor are prominent gradients apparent across the deepest part of the footprint. Greater caution is nevertheless required near the lower-coverage boundaries, and visual inspection alone cannot exclude subtle background offsets. The profiles presented in Sect.~\ref{sec:profiles} provide a more sensitive assessment of possible large-scale oversubtraction. Apart from the localized artifacts described above, the mosaics therefore appear sufficiently reliable for analyzing coherent diffuse emission on the spatial scales considered here.

\begin{figure*}
    \centering
    \includegraphics[width=0.7\textwidth]{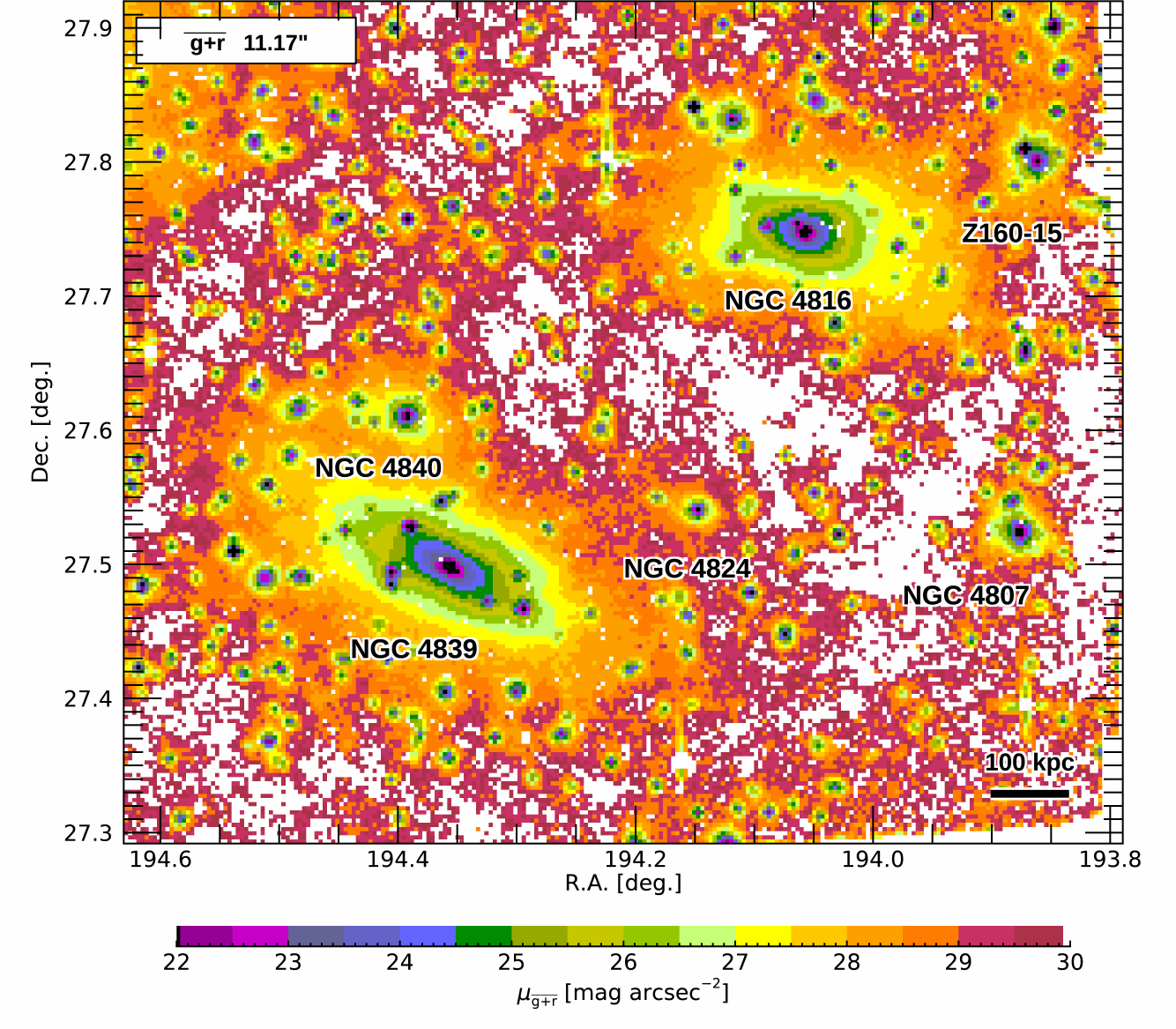}
    \caption{Low-surface-brightness structures in the NGC~4839--NGC~4816 region. The panel shows the masked mean $g+r$ image spatially binned to a pixel scale of 11.14~arcsec~pixel$^{-1}$. The color scale indicates surface brightness in units of mag~arcsec$^{-2}$. The main galaxies and diffuse structures discussed in the text are labeled.}
    \label{fig:Puente}
\end{figure*}

\subsection{Diffuse structures in the NGC~4839--NGC~4816 complex}
\label{sec:4839_4816}

Figure~\ref{fig:Puente} shows the region surrounding NGC~4839 and NGC~4816 in the southwestern part of Coma. The surface-brightness map was constructed using a masking and spatial-binning procedure similar to that adopted for Fig.~\ref{fig:Coma_mag}, but with a $10\times10$ binning of the native images, corresponding to a pixel scale of 11.14~arcsec~pixel$^{-1}$. Following the procedure described in Sect.~\ref{sec:global_view}, the nominal $3\sigma$ limits at the $11.14\times11.14~\mathrm{arcsec}^{2}$ binned-pixel scale are $\mu_{\mathrm{lim},g}=30.2~\mathrm{mag\,arcsec^{-2}}$ and $\mu_{\mathrm{lim},r}=29.7~\mathrm{mag\,arcsec^{-2}}$ at the maximum local exposure time. Propagating the band-specific fluctuations for the arithmetic mean of the two mosaics yields an effective nominal limit of $\mu_{\mathrm{lim},\overline{g+r}}=30.3~\mathrm{mag\,arcsec^{-2}}$. In the deepest regions of Fig.~\ref{fig:Puente}, spatially coherent emission can be readily traced to $\mu_{\overline{g+r}}\simeq30.0~\mathrm{mag\,arcsec^{-2}}$. Toward the southwestern boundary, the shorter local integration time reduces the sensitivity and produces a visibly noisier background.

The field is dominated by the extended stellar envelopes of NGC~4839 and NGC~4816, two massive early-type galaxies separated by approximately 600~kpc in projection. NGC~4839, the central galaxy of an infalling group that has likely already crossed the cluster core \citep{Lyskova2019, Churazov2021, Bonafede2021, Lal2022,Oh2023, JimenezTeja2025}, exhibits a broad, slightly asymmetric LSB envelope. By contrast, the diffuse envelope of NGC~4816 remains more clearly detached from both the central ICL and the emission associated with NGC~4839, consistent with an earlier stage of interaction with Coma \citep{Oh2023, JimenezTeja2025}. Several other bright galaxies, including NGC~4840, NGC~4824, and NGC~4807, are embedded in the surrounding network of faint diffuse emission.

The most prominent intergalactic diffuse structure in the field is the extremely faint bridge-like feature extending between the outer regions of NGC~4839 and NGC~4816, previously reported by \citet{JimenezTeja2025}. After masking superimposed sources, the bridge has an estimated surface brightness of $\mu_{\overline{g+r}}\sim29$--$30~\mathrm{mag\,arcsec^{-2}}$. It has an approximate projected width of 100~kpc and can be traced over a length of about 500~kpc. Its surface brightness and projected length are comparable to those of the Giant Coma Stream \citep{Roman2023}, although the bridge is substantially broader and less sharply defined. A detailed investigation of its origin is deferred to a forthcoming paper.

Additional LSB features are visible throughout the complex. NGC~4824 appears embedded in diffuse emission and may be connected to the outer envelope of NGC~4839 by a faint stellar filament with a characteristic surface brightness of $\mu_{\overline{g+r}}\sim29~\mathrm{mag\,arcsec^{-2}}$. Toward the northern part of the field, Z160-15 exhibits an extended, approximately north--south envelope reaching a similar surface brightness. This structure does not appear to connect directly with NGC~4816 and may instead trace a local tidal distortion or a response to the broader group or cluster tidal field. Near the southwestern boundary, NGC~4807 shows a possible north--south extension. Given the reduced local integration time and visibly noisier background, we regard this feature as tentative.
 
\subsection{Photometric profiles of NGC~4839 and NGC~4816}
\label{sec:profiles}

\begin{figure*}
    \centering
    \includegraphics[width=\textwidth]{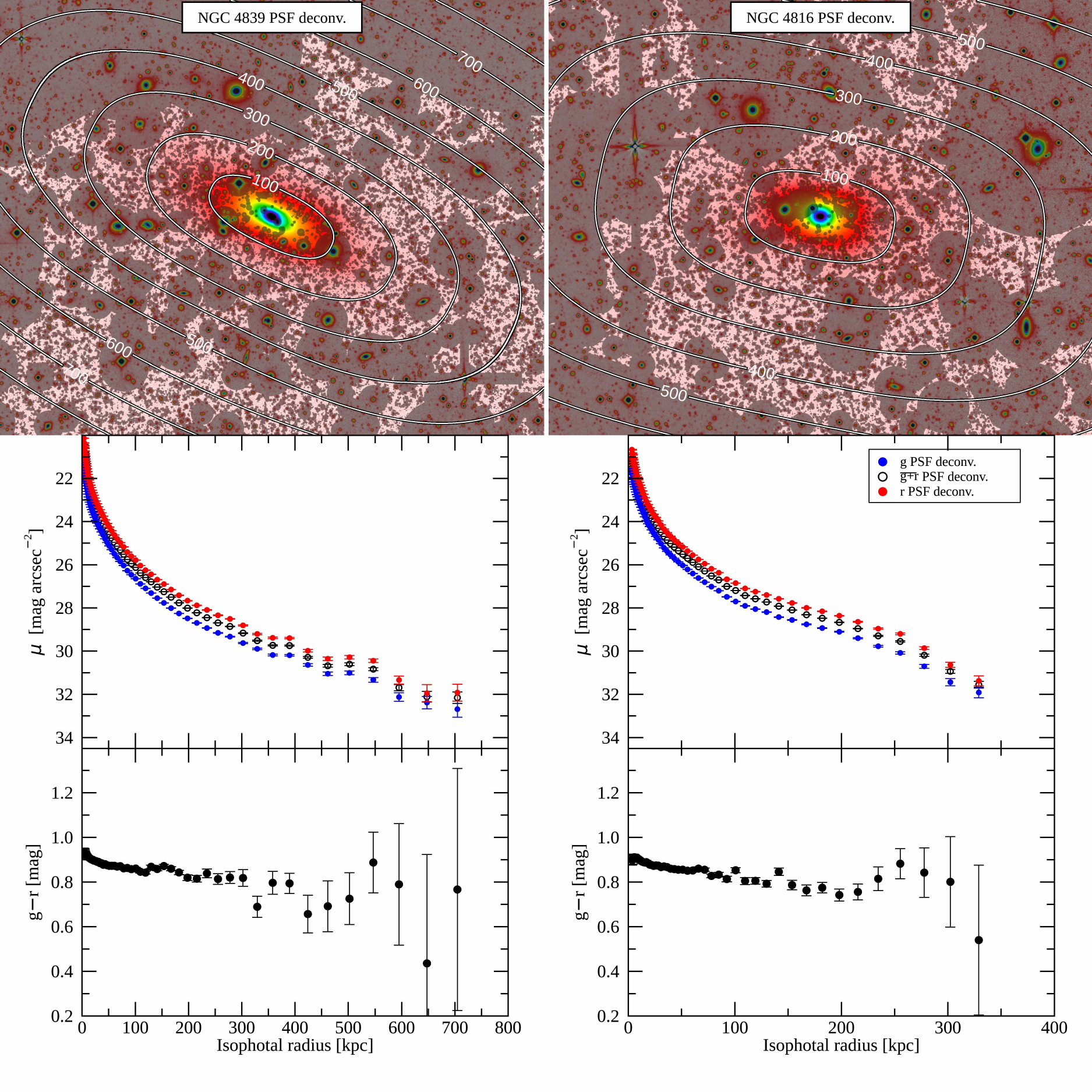}
    \caption{Photometric profiles of NGC~4839 (left) and NGC~4816 (right). The upper panels show the PSF-deconvolved $\overline{g+r}$ images together with the isophotal-radius contours adopted for the profile extraction. The middle panels present the surface brightness profiles measured in the $g+r$, $g$, and $r$ images, while the lower panels show the corresponding $g-r$ color profiles.}
    \label{fig:Perfiles}
\end{figure*}

Figure~\ref{fig:Perfiles} presents the surface-brightness and $g-r$ color profiles of NGC~4839 and NGC~4816. The profiles were extracted from the PSF-deconvolved $g+r$, $g$, and $r$ images. Because both galaxies exhibit strongly boxy morphologies, their profiles were extracted using customized isophotal apertures, which are better suited to following their light distributions than conventional circular or elliptical apertures. The construction of the isophotal-radius maps, masking strategy, PSF-deconvolution procedure, profile extraction, and uncertainty estimation are described in Appendix~\ref{app:photometric_profiles_4839_4816}.

The surface-brightness profiles reveal markedly different outer-envelope structures. NGC~4839 can be traced to an isophotal radius of approximately 700~kpc, where the mean combined surface brightness reaches $\mu_{\overline{g+r}}\sim32$--$32.5~\mathrm{mag\,arcsec^{-2}}$. Its profile shows a broadly smooth decline over most of the measured radial range, with no obvious evidence of a pronounced truncation. By contrast, NGC~4816 is detected to a maximum isophotal radius of approximately 330~kpc, where the combined profile approaches $\mu_{\overline{g+r}}\sim32~\mathrm{mag\,arcsec^{-2}}$. Its surface brightness decreases smoothly out to approximately 250~kpc before steepening markedly, indicating a more spatially confined outer envelope than that of NGC~4839.

For context, \citet{Janowiecki2010} derived $V$-band surface-brightness profiles for five luminous Virgo ellipticals spanning an estimated stellar-mass range of $M_{\star}\simeq(1.4$--$6.4)\times10^{11},M_{\odot}$ \citep{Cappellari2011,Ma2014}, typically tracing their outer isophotes to projected radii of approximately 100--150~kpc and surface-brightness levels of $\mu_V\sim28$--$29~\mathrm{mag\,arcsec^{-2}}$, illustrating the strong dependence of the measured extent on profile depth. A more extreme comparison is provided by the cD galaxy IC~1101 in Abell~2029 with an estimated stellar mass of $M_{\star}\simeq1.1\times10^{12}\,M_{\odot}$ \citep{Dullo2010}, whose diffuse $R$-band envelope was traced to an elliptical radius of $d\simeq425\,h^{-1}$~kpc, corresponding to approximately 610~kpc for $h=0.7$ \citep{Uson1991}. More recent ultra-deep imaging identifies the edge of its main stellar body at $R_{\mathrm{edge}}=260\pm38$~kpc, while tracing the extended BCG+ICL component to approximately 475~kpc and coherent diffuse emission to approximately 620~kpc \citep{Marrero2026}. Although the passbands, measurement geometries, definitions of radius, and surface-brightness depths differ, the isophotal extent measured around NGC~4839 is comparable to that of the exceptionally extended diffuse-light distribution surrounding IC~1101. At such large radii, however, the measured extent is naturally sensitive to the depth of the profiles.

NGC~4839 and NGC~4816 exhibit broadly similar inner colors, with central values of approximately $g-r\simeq0.9$~mag and progressively bluer colors toward intermediate radii. Negative color gradients are commonly observed in massive early-type galaxies \citep[e.g.,][]{Roche2010, LaBarbera2012, Tal2011, DSouza2014}, and similar outward bluing has been traced into the extended halos of M87, M49, and NGC~5128 \citep{Rudick2010, Mihos2013, Rejkuba2014}. At large radii, stacked color profiles of massive early-type galaxies generally tend to flatten \citep{Tal2011, DSouza2014}. NGC~4839 appears to show a similar behavior, although the uncertainties increase substantially in its outermost bins. NGC~4816 instead exhibits a more clearly defined change in its color profile: after becoming progressively bluer out to approximately 200~kpc, it turns redward and reaches $g-r\simeq0.85$--$0.9$~mag at approximately 250--280~kpc. This redward turn approximately coincides with the region in which the surface-brightness profile begins to steepen. In M49, an outer stellar shell has been proposed to produce a similar behavior to that of NGC~4816 \citep{Mihos2013}, demonstrating that accreted substructure can produce localized redward color deviations in photometric profiles. A similar contribution could plausibly affect the outer profile of NGC~4816.

The profiles presented here provide the first quantitative characterization of the extended stellar envelopes of NGC~4839 and NGC~4816, highlighting their markedly different radial extents and outer color behaviors. A forthcoming paper will examine the origin and evolutionary implications of these structures in greater detail.

\section{Discussion}
\label{sec:discussion}

In this article, we introduce the HERON Coma Project, a deep $g$- and $r$-band imaging campaign carried out with the LSB-optimized 71-cm Jeanne Rich Telescope to trace the faint stellar record of the Coma complex and ongoing assembly. As a presentation paper of the project, this work focuses on assessing the reliability and LSB performance of the dataset, while presenting an overview of the results.

The observational and data-processing strategy was designed to maximize sensitivity in the LSB regime. The empirical relation between integration time and nominal limiting surface brightness, determined from the background fluctuations using the standard metric of $3\sigma$ in $10\times10~\mathrm{arcsec}^{2}$ apertures \citep[e.g.,][]{Roman2020}, follows the approximately $t_{\mathrm{exp}}^{-1/2}$ behavior expected for random-noise-dominated observations (Fig.~\ref{fig:Depth_plot}). The fitted coefficients of 2.54 are close to the value of 2.5 expected from ideal random-noise scaling in the adopted parametrization (Eqs.~\ref{eq:depth_g} and \ref{eq:depth_r}), indicating the efficiency of the observing and data-processing strategy.

A surface-brightness limit derived from background fluctuations should not be interpreted as an absolute detection threshold. Instead, it represents the nominal depth attainable when systematic effects do not impose a brighter practical floor. Photometry of extended LSB emission may become limited at brighter levels by several systematics \citep[e.g.,][]{Mihos2019, Kelvin2023, Watkins2024}. The analyses conducted here provide a direct test of whether these effects set the practical sensitivity of the current dataset or whether the measurements remain predominantly background-noise limited over the spatial scales examined.

The surface-brightness levels reached in the enhanced images can be compared directly with the random-noise sensitivity expected at their respective spatial scales. For the $5.57\times5.57~\mathrm{arcsec}^{2}$ binned pixels adopted in Fig.~\ref{fig:Coma_mag}, the effective nominal $3\sigma$ limit in the region of maximum coverage is $\mu_{\mathrm{lim},\overline{g+r}}=29.54~\mathrm{mag\,arcsec^{-2}}$, closely matching the level of $\mu_{\overline{g+r}}\simeq29.5~\mathrm{mag\,arcsec^{-2}}$ to which spatially coherent emission can be visually traced. For the $11.14\times11.14~\mathrm{arcsec}^{2}$ binned pixels adopted in Fig.~\ref{fig:Puente}, the larger averaging area yields an effective nominal $3\sigma$ limit of $\mu_{\mathrm{lim},\overline{g+r}}=30.29~\mathrm{mag\,arcsec^{-2}}$. Coherent diffuse emission is clearly detectable up to $\mu_{\overline{g+r}}\simeq30.0~\mathrm{mag\,arcsec^{-2}}$, with the modest difference from the maximum nominal depth being consistent with the lower and less homogeneous exposure time across the southwestern footprint. 

The PSF-deconvolved photometric profiles reach lower surface brightness by averaging over large isophotal annuli rather than fixed image bins. For NGC~4839, emission is measured at $\mu_{\overline{g+r}}=32.16~\mathrm{mag\,arcsec^{-2}}$ at $r_{\mathrm{isophotal}}=705$~kpc with a data-derived signal-to-noise ratio of 4.1. The effective weighted area of the annulus after masking and clipping is $2.62\times10^{4}~\mathrm{arcsec}^{2}$, and the measured uncertainty corresponds to a formal $3\sigma$ sensitivity of $\mu_{\mathrm{lim},\overline{g+r}}=32.50~\mathrm{mag\,arcsec^{-2}}$. For NGC~4816, the outermost measured point at $r_{\mathrm{isophotal}}=329$~kpc has $\mu_{\overline{g+r}}=31.56~\mathrm{mag\,arcsec^{-2}}$ and a data-derived signal-to-noise ratio of 7.0. Its effective weighted area is $1.51\times10^{4}~\mathrm{arcsec}^{2}$, yielding a formal $3\sigma$ sensitivity of $\mu_{\mathrm{lim},\overline{g+r}}=32.48~\mathrm{mag\,arcsec^{-2}}$. The profiles are shown only to the last bins before an outer upturn becomes apparent, beyond which the measured emission can no longer be associated unambiguously with either galaxy. These upturns are most naturally explained by overlap with the surrounding ICL and other diffuse structures, which will be studied in forthcoming results, although spatially varying background residuals cannot be excluded. The still appreciable signal-to-noise ratios indicate that, if the galaxies were isolated and their backgrounds could be constrained independently, both profiles could in principle be traced closer to their formal $3\sigma$ limits of approximately $32.5~\mathrm{mag\,arcsec^{-2}}$.

Although truncating the profiles at the onset of these upturns prevents their behavior at larger radii from being tested, the absence of abrupt downturns up to that point provides a complementary, albeit inconclusive, test for significant background oversubtraction \citep[e.g.,][]{Kelvin2023, Watkins2024}. This is particularly clear for NGC~4839, whose profile follows a steady decline to $\mu_{\overline{g+r}}=32.2~\mathrm{mag\,arcsec^{-2}}$. NGC~4816 displays an apparent truncation around $\mu_{\overline{g+r}}\approx29.5~\mathrm{mag\,arcsec^{-2}}$, but its morphology and color profile behavior, at a relatively bright surface brightness, suggest that this feature may have a physical origin. This feature will be investigated in a separate study. The profiles therefore show no clear evidence of substantial oversubtraction within the radial and surface-brightness ranges probed here, although subtler biases cannot be excluded as the measurements approach $32$--$32.5~\mathrm{mag\,arcsec^{-2}}$. This conclusion should not be extrapolated directly to the ICL, whose angular extent is comparable to the field of view of the instrument. Its separation from the background is consequently more challenging, and genuine diffuse emission may be partially absorbed into the fitted sky model.

The consistency between the surface-brightness levels reached in our analyses and the nominal depths predicted from background fluctuations provides no clear indication of a systematics-dominated detection floor over the spatial scales examined here. Low-level residual systematics remain present, particularly localized ghost-like artifacts from off-axis stars described in Sect.~\ref{sec:global_view}, but they do not appear to determine the global sensitivity of the current measurements. Within the scope of these analyses and wherever sufficiently uncontaminated spatial coverage is available, the HERON Coma dataset therefore appears to be predominantly limited by background noise, supporting its reliability for studies of faint extended structures.

Under this working interpretation, the empirical relation shown in Fig.~\ref{fig:Depth_plot} provides a quantitative estimate of the integration required to reach still fainter nominal depths. Propagating the band-specific relations and adopting the standard $3\sigma$ metric in $10\times10~\mathrm{arcsec}^{2}$ apertures, nominal limits of $\mu_{\mathrm{lim},\overline{g+r}}=30.5$, $31.0$, and $31.5~\mathrm{mag\,arcsec^{-2}}$ would require approximately 65, 160, and 400~h of accepted exposure per band, respectively. At these depths, however, systematic effects are likely to become increasingly important and may impose a brighter practical detection floor than predicted by random-noise scaling.

Source crowding is among the first limitations expected to become increasingly important as the imaging depth increases \citep{vanDokkum2020}. Faint compact sources convolved with the relatively broad $3.0$--$3.6$~arcsec FWHM of the HERON data may blend into an apparently homogeneous unresolved background. We argue that this systematic does not dominate the present analysis because Coma lies close to the north Galactic pole, where the density of foreground Milky Way stars is low. Moreover, although its central FWHM is relatively poor, the HERON PSF performs well over scales of several arcsec to 1 arcmin (Appendix~\ref{sec:PSF}; Fig.~\ref{fig:PSFs_comparison}). Crowding is nevertheless already apparent in the HERON data: reliable photometry of the much narrower Giant Coma Stream required deeper WHT follow-up with better seeing to separate its diffuse emission from superposed sources \citep{Roman2023}. At still fainter levels, improved angular resolution may therefore become essential to prevent resolved and unresolved-source fluctuations from imposing a practical detection floor. FWHM values of approximately 1~arcsec or below, achievable from ground-based observations, would reduce the effective resolution-element area by about an order of magnitude relative to HERON, substantially mitigating source blending and crowding.

Although the extended PSF is expected to vary as optical cleanliness and observing conditions evolve, this variability has not represented a major limitation in this project. The favorable low-scatter properties of its PSF allowed the extended halos of bright stars to be subtracted relatively straightforwardly, with only localized residuals around a few particularly bright sources (Sect.~\ref{sec:global_view}). Provided that a similarly high-quality PSF is maintained, scattered-light contamination should remain a controllable systematic at greater depths. Its impact could be reduced further by obtaining repeated PSF calibration sequences throughout the science observations and constructing temporally matched models, following approaches such as that of \citet{Liu2022}. This requirement is not exclusive to ground-based imaging: although space observatories avoid atmospheric variability, their optical state can also evolve, as illustrated by the gradual throughput loss caused by water-ice accumulation on the Euclid optics \citep{Schirmer2023}. While this example does not directly demonstrate a change in the extended PSF, it shows that continuous monitoring of optical performance is also required from space.

The HERON observations have benefited particularly from the strong suppression of internal reflections provided by the telescope’s simple, LSB-optimized optical design \citep{Rich2019}. The remaining contamination is limited mainly to scarce ghost-like features generated by bright stars outside or near the edge of the field of view, which are present in only a subset of the individual exposures. Their position relative to the astronomical target generally changes across the wide dithering pattern, causing most of their signal to be attenuated during coaddition. Detectable residuals persist only where similar telescope pointings were repeated sufficiently often for the corresponding ghosts to overlap at approximately the same sky position. These residuals locally affect the faintest emission, as discussed in Sect.~\ref{sec:global_view}, but occupy restricted regions and do not impose a global detection floor. We regard the suppression of internal reflections as an essential instrumental requirement for projects seeking reliable measurements not only at fainter surface-brightness levels, but even at depths comparable to those reached here. The source- and field-dependent morphology of optical reflections makes them exceptionally difficult to model, while their subtraction rarely produces completely clean residuals, potentially having a major impact on datasets constructed from large numbers of exposures and large total integration times \citep{Slater2009, Karabal2017}.

As discussed above, extending the nominal surface-brightness depth with a HERON-like 71-cm system rapidly becomes prohibitive in terms of total integration time. In the ideal background-noise-dominated regime, and assuming unchanged throughput, sky brightness, angular sampling, observing efficiency, and instrumental performance, the required exposure scales inversely with collecting area as $t(D)=t_{\mathrm{HERON}}(0.71~\mathrm{m}/D)^2$. The nominal $3\sigma$ limits of $\mu_{\mathrm{lim},\overline{g+r}}=30.5$, $31.0$, and $31.5~\mathrm{mag\,arcsec^{-2}}$ in $10\times10~\mathrm{arcsec}^{2}$ apertures would therefore require approximately 33, 81, and 202~h of accepted exposure per band with a 1-m telescope; 5.2, 12.9, and 32.3~h with a 2.5-m telescope; and 1.3, 3.2, and 8.1~h with a 5-m telescope. 

In practice, however, increasing the aperture while preserving a comparable or wider field of view often requires a more complex optical system that can introduce scattered light and internal reflections. These features may establish a systematic floor that cannot be overcome through longer integration. A detailed assessment of possible telescope designs is beyond the scope of this work, but nevertheless, the simple scaling exercise above shows that substantially fainter depths are observationally feasible if larger-aperture instrumentation can be developed while preserving the systematic-control requirements identified here. An LSB-optimized telescope combining a larger collecting area with reflection-suppressed optics, a wide field of view, good PSF quality, and a site delivering image FWHM of approximately 1~arcsec or better would provide a realistic path beyond the current observational frontiers of the LSB Universe.

\section{Summary and conclusions}

We have presented the HERON Coma Project, a deep $g$- and $r$-band imaging campaign on the Coma cluster carried out with the LSB-optimized 71-cm Jeanne Rich Telescope to provide an unprecedented view of the faint extended emission throughout the Coma cluster. The scientific potential of the dataset has already been demonstrated by the discovery of the Giant Coma Stream, the faintest stellar stream detected by surface photometry ($\mu_{g,max}$~$=$~29.5~mag~arcsec$^{-2}$) \citep{Roman2023} and a dedicated analysis of the extended ICL of Coma \citep{JimenezTeja2025}. As a representative application of the further science enabled by the dataset, we examine the NGC~4839--NGC~4816 region, where spatially coherent emission is traced to $\mu_{\overline{g+r}}\simeq30~\mathrm{mag\,arcsec^{-2}}$ at a spatial scale of 11.14~arcsec. The PSF-deconvolved profiles of NGC~4839 and NGC~4816 extend to isophotal radii of 705 and 329~kpc, reaching $\mu_{\overline{g+r}}=32.16$ and $31.56~\mathrm{mag\,arcsec^{-2}}$ at signal-to-noise ratios of 4.1 and 7.0, respectively. The observed improvement in sensitivity with exposure time, together with the agreement between nominal background-noise limits, direct detections, and aperture-averaged photometric profiles across multiple spatial scales, provides no clear evidence of a systematics-dominated detection floor within the regimes explored here, although localized artifacts remain. These results demonstrate the ability of modest ground-based telescopes with low-scatter optics and carefully controlled systematics to provide highly competitive measurements of faint extended structures. The calibrated mosaics are publicly released. Several additional studies exploiting different aspects of these data are currently in preparation, and the scientific exploitation of the HERON Coma Project will continue through a series of forthcoming papers.

\section*{Data availability}
The final calibrated $g$- and $r$-band mosaics, together with the corresponding weight maps and ancillary products, are publicly available at the CDS via anonymous ftp to
\url{cdsarc.u-strasbg.fr}. Pending completion of the CDS ingestion process, the data products can also be downloaded from the HERON Coma Project website at \url{https://jromanastro.wordpress.com/the-heron-coma-cluster-project/}. Additional data products are available from the corresponding author upon request.

\begin{acknowledgements}
We thank the referee for a careful and constructive analysis of our work. We thank Qing Liu and the Dragonfly collaboration for providing us with the Dragonfly PSFs used in this work. We thank Ignacio Trujillo and Mireia Montes for interesting discussions about the results. We acknowledge the late Peter Erwin for the development and maintenance of \texttt{IMFIT}, which was used in this work and many others. JR acknowledges financial support from Plan Propio de Investigación 2025 submodalidad 2.3 of the University of Córdoba. PMSA acknowledges that part of this research was sponsored by the National Aeronautics and Space Administration (NASA) through a contract with ORAU. The views and conclusions contained in this document are those of the authors and should not be interpreted as representing the official policies, either expressed or implied, of NASA or the U.S. Government. The U.S. Government is authorized to reproduce and distribute reprints for Government purposes notwithstanding any copyright notation herein.
\end{acknowledgements}

\bibliographystyle{aa}
\bibliography{Refs}

\appendix
\nolinenumbers

\section{Extended point spread function characterization}
\label{sec:PSF}

Accurate characterization of the extended PSF is essential for LSB studies. Its wings can contaminate diffuse emission and bias sky estimates \citep{Slater2009,InfanteSainz2020,Watkins2024}, while redistributing galaxy light to large radii and distorting surface-brightness and color profiles \citep{Sandin2014,Sandin2015,TrujilloFliri2016,Karabal2017}.

The dedicated PSF observations were concentrated toward the end of each observing campaign and consist of individual 300-s exposures of bright stars. Targets were selected among the brightest stars observable from the site, while avoiding fields strongly affected by Galactic cirrus, thereby maximizing the signal-to-noise ratio in the extended PSF wings. As in the Coma observations, each target was observed using a wide dithering pattern. This strategy displaced field-dependent and localized ghost features relative to the stellar centroid, preventing them from adding coherently when the star-centered exposures were combined. Approximately 50 bright-star exposures were obtained in each band, distributed among about eight stars per filter.

The images were processed using the same basic reduction steps and the corresponding science superflats described in Sect.~\ref{sec:reduction}. Each exposure was background-subtracted, resampled onto a common grid centered on the stellar centroid, and aggressively masked to remove unrelated sources. The images were then normalized to a common stellar flux using circular-aperture photometry, following procedures similar to those described by \citet{Roman2020} and \citet{InfanteSainz2020}. After centering, masking, and flux normalization, the exposures were combined using a $3\sigma$-clipped mean.

No single stellar sample provides an unsaturated, high-signal-to-noise measurement of the PSF over its full radial extent. The brightest stars provide the strongest constraints on the faint outer wings, but their central regions are heavily saturated. We therefore reconstructed the intermediate radial range using stars of intermediate apparent brightness observed in additional HERON fields contemporaneously with the Coma campaign. Although these stars remained saturated in their innermost regions, they provided the radial coverage required to connect the unsaturated cores measured from fainter stars with the extended wings constrained by the brightest targets.

The central PSF core was constructed from a large sample of unsaturated stars selected from the final Coma mosaics and therefore reproduces the effective FWHM of the coadded Coma observations. The complete PSF was assembled from the radial ranges constrained by the bright, intermediate-brightness, and unsaturated stellar samples, using each sample only where it remained unsaturated and provided adequate signal-to-noise. The profiles were scaled and matched over their overlapping radial intervals before being combined. Comparable multi-brightness approaches have been used to construct extended PSFs over a wide dynamic range in other deep-imaging studies \citep{Roman2020, InfanteSainz2020, GarateNunez2024, Sedighi2025}.

\begin{figure}
    \centering
    \includegraphics[width=\columnwidth]{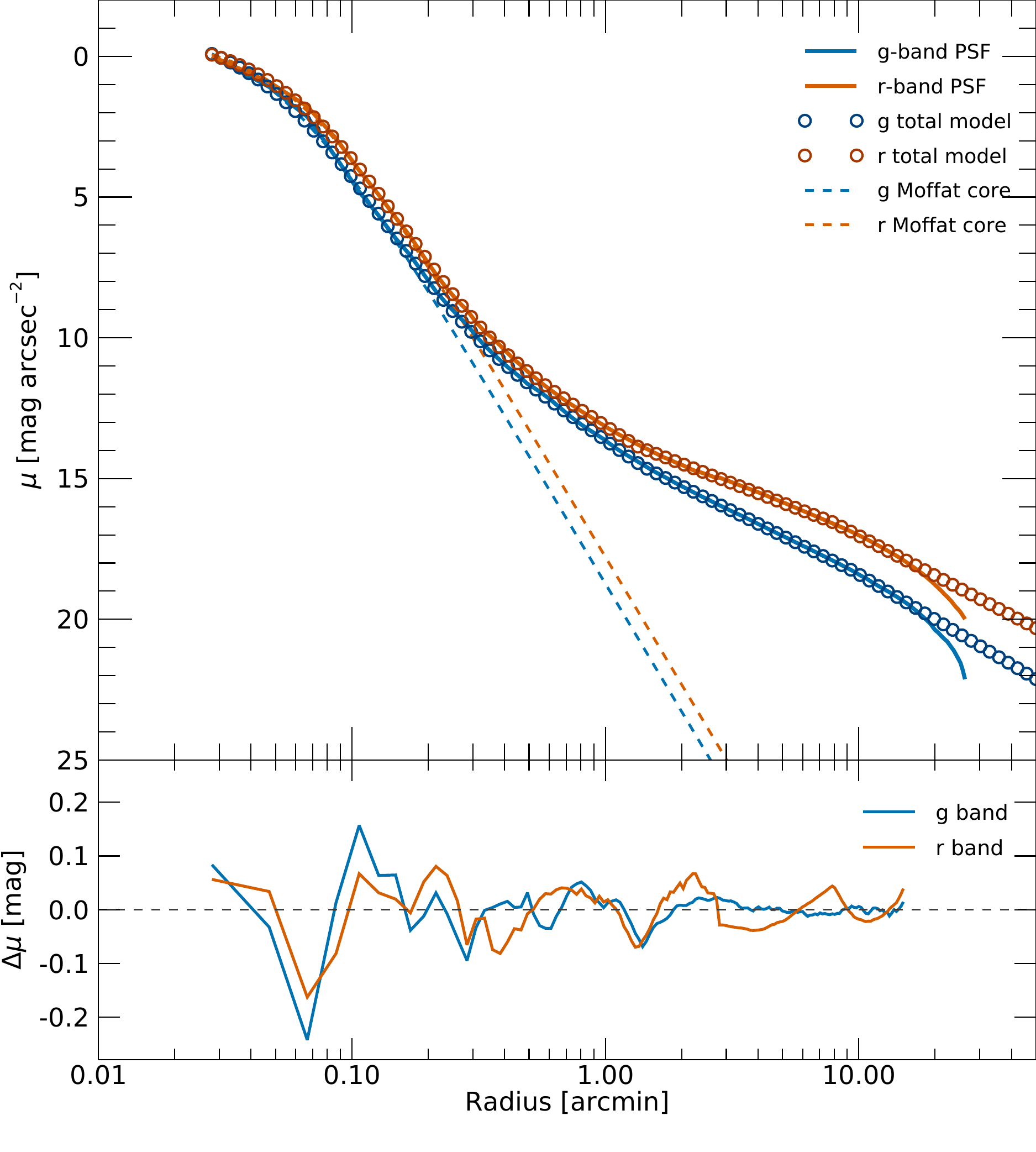}
    \caption{Azimuthally averaged HERON $g$- and $r$-band PSF profiles and their simultaneous Moffat-core plus multi-power-law aureole fits. Each observed profile is normalized to a central surface brightness of $\mu=$~0~mag~arcsec$^{-2}$. Solid curves show the empirical profiles, open circles show the best-fitting total models, and dashed curves show the Moffat components. Beyond the 15-arcmin fitting interval, the open-circle continuation represents an extrapolation of the final aureole components. The lower panel presents the data-minus-model residuals over the fitted radial range.}
    \label{fig:PSF_profiles}
\end{figure}

To improve the signal-to-noise ratio in the faint outer wings, we derived an azimuthally averaged radial profile from the combined PSF image in each band and used it to construct a circularly symmetric two-dimensional model. This approach is also motivated by the spatial variation in the diffraction-spike pattern, whose geometry and relative intensity change with detector position. A two-dimensional PSF constructed by stacking stars observed at different positions would smear these non-axisymmetric features and would not reproduce the diffraction pattern of any individual star. Its application could consequently leave mismatched positive and negative residuals around the spikes. Azimuthal averaging instead provides a stable description of the approximately axisymmetric stellar halo, which is the component that can be transferred most reliably across the field, while the position-dependent spikes must be treated through masking. Their flux is not, however, entirely absent from the radial profile. Because the stellar images were combined using a $3\sigma$-clipped mean, the contribution of the spikes depends on the fraction of each annulus that they occupy. At large radii, the symmetric halo covers a much larger area and dominates the recovered profile. Closer to the center, the spikes occupy a non-negligible fraction of the sampled area, allowing some of their flux to enter the azimuthal average even though their morphology cannot be reproduced by the circular model. The resulting HERON PSF models extend to approximately 25~arcmin, and their surface-brightness profiles are shown in Fig.~\ref{fig:PSF_profiles}.

Following an approach similar to that of \citet{Liu2022}, we modeled each band-specific PSF as the sum of a Moffat core and a continuous, segmented power-law aureole. The full model was fitted simultaneously to all valid measurements out to 15~arcmin, with the aureole constrained from three times the adopted image FWHM, corresponding to 9.3~arcsec in $g$ and 10.5~arcsec in $r$. Models containing one, two, and three aureole components were compared using the Bayesian information criterion (BIC), requiring an improvement of at least 10 to accept an additional component. Each segment was required to contain at least ten radial measurements, and multiple initial solutions were explored to reduce sensitivity to local minima. Parameter uncertainties were estimated from 200 residual-bootstrap realizations. The resulting profiles, best-fitting models, and residuals are shown in Fig.~\ref{fig:PSF_profiles}.

The BIC favors three aureole components in both bands. Relative to the one-component models, the improvements are $\Delta\mathrm{BIC}=275$ in $g$ and 248 in $r$; relative to the two-component models, they are 65 and 153, respectively. All 200 bootstrap realizations converged. In the $g$ band, the fitted core has $\mathrm{FWHM}=3.023\pm0.043$~arcsec and $\beta=3.023\pm0.038$. The aureole has exponents $\alpha_1=2.513\pm0.022$, $\alpha_2=1.783\pm0.007$, and $\alpha_3=2.131\pm0.052$, separated by breaks at $1.412\pm0.026$ and $9.401\pm0.436$~arcmin. In the $r$ band, the fitted core has $\mathrm{FWHM}=3.640\pm0.055$~arcsec and $\beta=3.020\pm0.038$, while the aureole exponents are $\alpha_1=2.282\pm0.030$, $\alpha_2=1.394\pm0.009$, and $\alpha_3=1.880\pm0.037$, with breaks at $1.334\pm0.026$ and $7.931\pm0.278$~arcmin. The complete fits have rms residuals of 0.033 and 0.037~mag in $g$ and $r$, respectively. Beyond the 15-arcmin fitting limit, the empirical profiles decline much more rapidly in intensity than the extrapolated models. Because the PSF was assembled from stars spanning a range of brightness, the outer wings are not detected with comparable signal-to-noise in all exposures. This heterogeneous sensitivity, combined with the $3\sigma$-clipped mean, could artificially steepen the terminal profile, although the origin of the downturn remains uncertain. We therefore regard the extrapolated final components as more representative of the outer PSF than the terminal empirical measurements, while emphasizing their model-dependent nature.

The resulting models should be understood as representative, band-dependent PSFs rather than exact exposure-by-exposure descriptions. Although the fit residuals are comparable, the $r$-band PSF has a broader core and shallower aureole components than the $g$-band PSF, indicating a larger scattered-light contribution rather than a poorer mathematical fit. The extended PSF can vary with the state of the optical surfaces, as illustrated by the changes measured after cleaning the Dragonfly lenses \citep{Liu2022}. The $r$-band calibration observations obtained in late May and June 2020 also overlapped with several early-season wildfires in central and southern California\footnote{See the California Department of Forestry and Fire Protection (CAL FIRE) 2020 Incident Archive: \url{https://www.fire.ca.gov/incidents/2020}.}. Enhanced aerosol loading from regional smoke, changes in the optical surfaces, and temporal differences between the dedicated PSF and Coma observations remain plausible contributors to the broader $r$-band PSF, although this interpretation is necessarily speculative.

\begin{figure*}
    \centering
    \includegraphics[width=\textwidth]{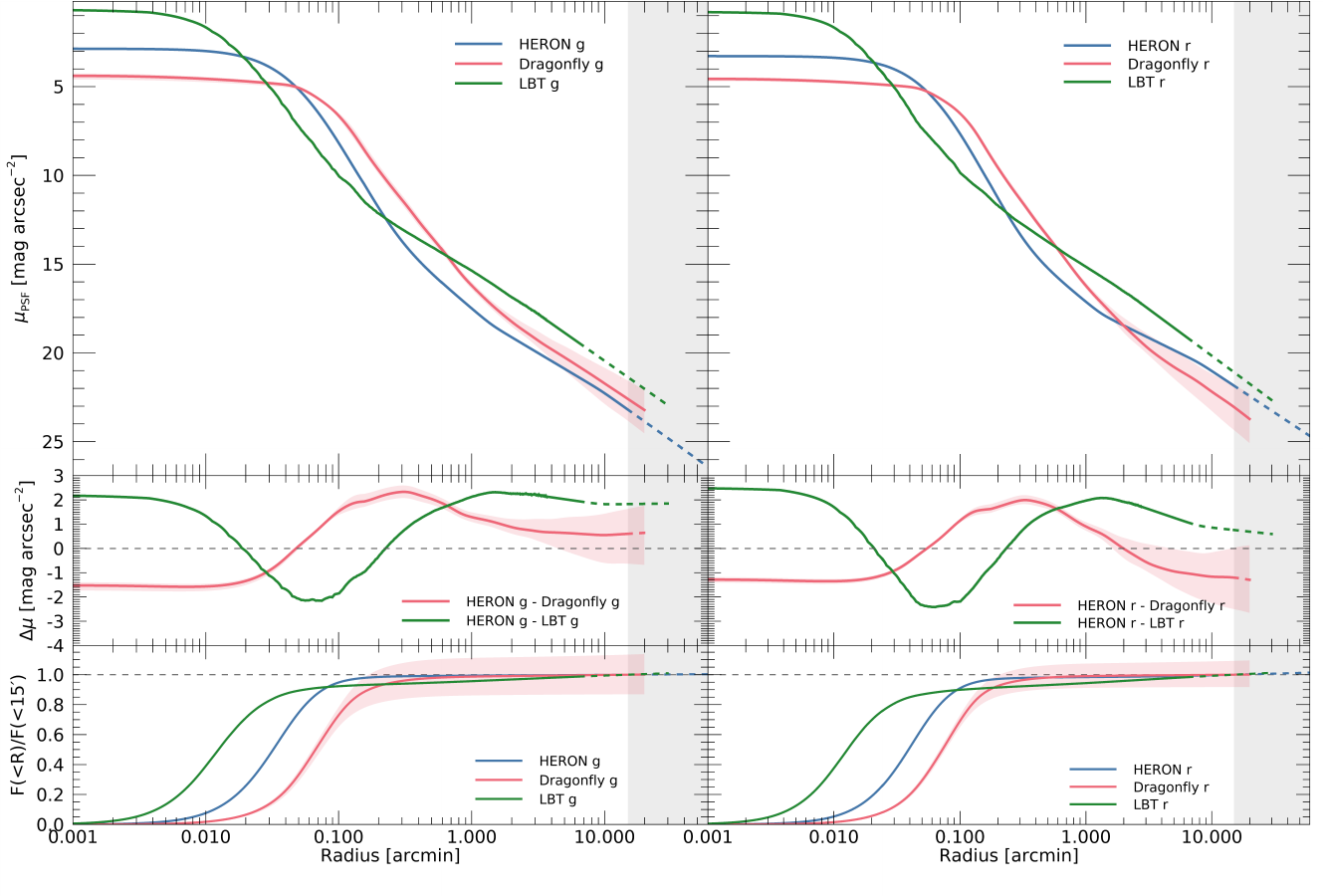}
    \caption{Comparison of the normalized HERON, Dragonfly, and LBT extended PSFs in the $g$ (left) and $r$ (right) bands. The upper panels show the radial surface-brightness profiles after normalizing each PSF to $F(<15~\mathrm{arcmin})=1$. The middle panels show $\Delta\mu=\mu_{\mathrm{HERON}}-\mu_{\mathrm{comparison}}$; positive values indicate that the HERON PSF is fainter at that radius. The lower panels show the enclosed-flux fraction $F(<R)/F(<15~\mathrm{arcmin})$. The Dragonfly $I_{50}$ profiles are shown by the central curves, with shaded regions spanning $I_{16}$--$I_{84}$ and using a mildly smoothed, shape-preserving representation described in the text. The HERON models are constrained by the fit out to 15~arcmin and are dashed beyond this radius, with the gray regions marking the extrapolated regime. The LBT profiles are solid over their empirically constrained radial ranges and dashed where the published extended models rely on extrapolation. Dragonfly is shown without extrapolation beyond the supplied 20-arcmin profiles.}
    \label{fig:PSFs_comparison}
\end{figure*}

We additionally compared the complete HERON profiles with two independent wide-angle PSF datasets, as shown in Fig.~\ref{fig:PSFs_comparison}. For the Dragonfly Telephoto Array, we use the PSF models constructed by \citet{Liu2022} for individual $1^\circ\times1^\circ$ tiles across a $100~\mathrm{deg}^{2}$ field. The comparison adopts the median profile, $I_{50}$, while the $I_{16}$--$I_{84}$ interval represents the spatial variation among tiles within the 16th and 84th quantiles. The models were derived from cirrus-subtracted images; although residual cirrus may affect individual tiles, the median should provide a stable representation of the field. We restrict the Dragonfly profiles to their supplied radial range and represent them on a common logarithmic radial grid using shape-preserving PCHIP interpolation in $\mu(\log r)$. For the Large Binocular Telescope (LBT), we use the circularized, publicly released extended two-dimensional g- and r-band PSF models derived from observations obtained with the Large Binocular Cameras \citep{Sedighi2025}. All profiles were placed on a common radial grid and normalized to unit enclosed flux within 15~arcmin. This normalization preserves their relative radial light distributions and enables direct comparison of local surface brightness and enclosed energy without requiring identical original stellar-flux scales. The resulting profiles and enclosed-flux distributions are shown in Fig.~\ref{fig:PSFs_comparison}.

The comparison shows why no single power-law exponent can provide an absolute ranking of wide-angle PSFs. The profiles cross at several radii, and their relative ordering depends on whether the core, intermediate aureole, or outer wings are considered. Within the common 15-arcmin normalization aperture, HERON encloses a larger fraction of its flux at small radii than the representative Dragonfly profiles in both bands, indicating a more concentrated radial light distribution over this range. Its local surface brightness is also fainter than the Dragonfly profiles over substantial subarcminute intervals, although this behavior is not maintained at every radius. The LBT profiles display a different balance between their compact cores and extended wings. Despite these differences, the profiles converge toward similar power-law slopes at large radii. The outermost HERON components have $\alpha_g=2.131\pm0.052$ and $\alpha_r=1.880\pm0.037$. Fits to the supplied Dragonfly profiles between 15 and 20~arcmin give $\alpha_g\simeq2.00$ and $\alpha_r\simeq2.20$, while the outer continuations of the LBT PSF models follow $\alpha\simeq2.10$ in both bands. Five of the six slopes therefore cluster within $\alpha\simeq2.00$--2.20, close to the classical $I(r)\propto r^{-2}$ behavior identified by \citet{King1971} and discussed by \citet{Sandin2014}. The HERON $r$-band value is the clear low-$\alpha$ outlier in this comparison, indicating a shallower decline and a stronger contribution from the outer wings. This result reinforces the concerns regarding the broader and more extended HERON $r$-band PSF discussed above.

\section{Bright-star subtraction}
\label{sec:starsubtraction}

\begin{figure*}
    \centering
    \includegraphics[width=\textwidth]{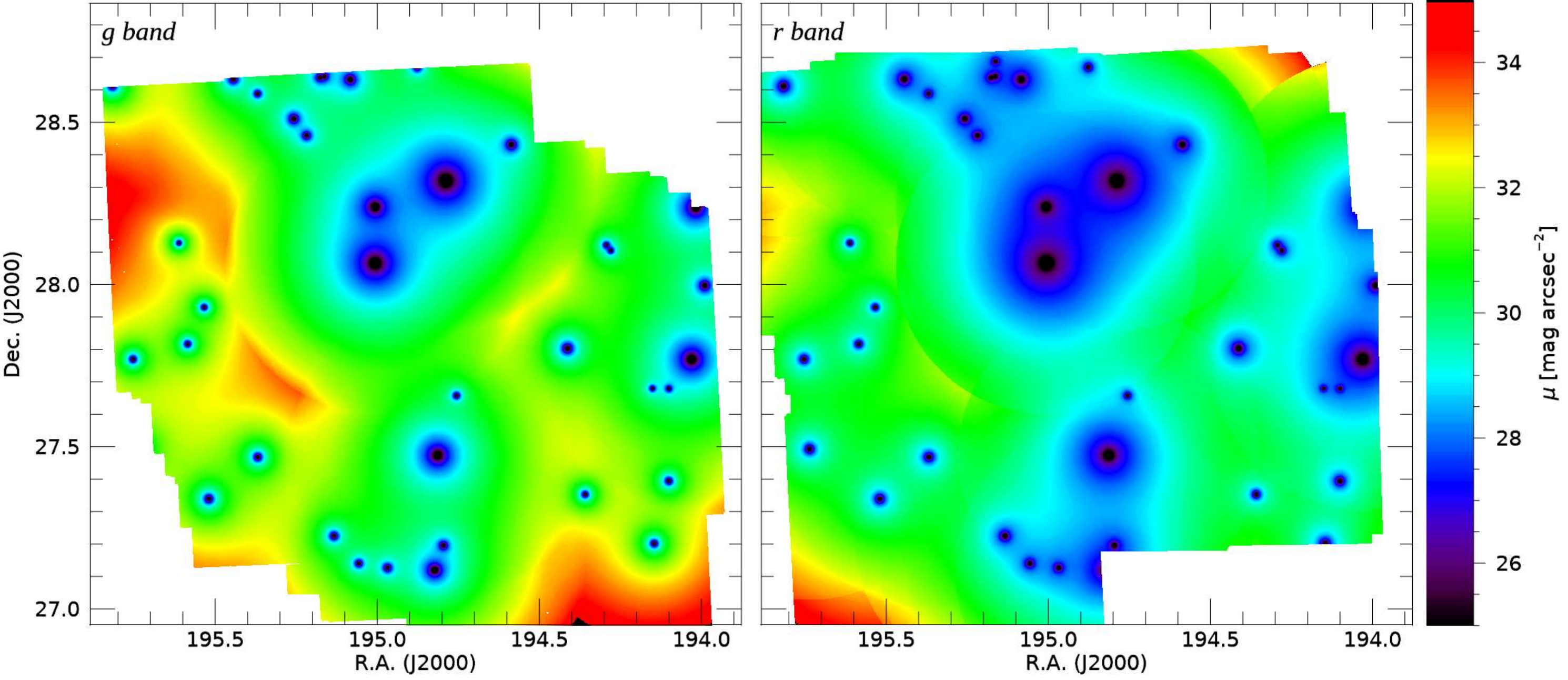}
    \caption{Modeled scattered-light contribution from bright stars across the HERON Coma footprint in the $g$ (left) and $r$ (right) bands. Each map was constructed by placing the corresponding extended PSF model at the fitted subpixel position of every selected star and scaling it to the measured stellar flux. The color scale indicates the surface brightness of the combined stellar model.}

    \label{fig:Mapas_SF}
\end{figure*}

To remove the dominant scattered-light contribution from the brightest stars within the Coma footprint, we fitted the centroid and flux normalization of each star using the unsaturated portions of its radial profile and the corresponding band-specific PSF model. The model was shifted to the fitted subpixel position and scaled to the measured stellar flux. We then constructed model images with the same dimensions, astrometric grid, and pixel scale as the corresponding Coma images and added the individual stellar models to obtain the total PSF-scattered-light distribution. These models were subtracted from the science exposures before the background-modeling and coaddition procedure described in Sect.~\ref{sec:coaddition}. Figure~\ref{fig:Mapas_SF} shows the combined bright-star models projected onto the final $g$- and $r$-band mosaic footprints.

The subtraction was restricted to 39 stars selected on the basis of their brightness and their impact on the low-surface-brightness images. Extending the procedure to fainter stars did not improve the results. At the surface-brightness levels where these sources remain detectable above the local background, their visible structure is dominated by non-axisymmetric diffraction spikes rather than by the approximately circular stellar halo. Applying a circular PSF model would consequently remove flux azimuthally from regions not occupied by the spikes, producing negative residuals while leaving most of the diffraction pattern unsubtracted. Fainter stars were therefore not modeled individually and their cores and diffraction spikes were instead included in the masks used for scientific analysis.

The subtraction is primarily intended to remove the extended, approximately axisymmetric halos of the selected bright stars. Some instruments exhibit position-dependent internal reflections that require explicit modeling \citep[e.g.,][]{Slater2009, Karabal2017}. The Jeanne Rich Telescope was specifically optimized for LSB imaging, with a simple prime-focus optical configuration and dedicated baffling designed to suppress internal reflections \citep{Rich2019}. No separate templates were therefore fitted for these features.

The wide dithering pattern further attenuates detector- and field-position-dependent ghost-like features by displacing them relative to the astronomical coordinate system between exposures. Such features are consequently diluted during coaddition. The circular PSF model therefore captures the dominant spatially transferable component of the stellar scattered light, although it cannot reproduce the complete two-dimensional pattern around every star. Residual diffraction spikes, faint ghost features, and local mismatches caused by temporal or field-dependent PSF variations may remain after subtraction. These residuals, together with the unsubtracted structures associated with fainter stars, were masked and excluded from the background modeling and scientific analysis.

\section{Spatial coverage and nominal-depth characterization}
\label{sec:depth_appendix}

\begin{figure*}
    \centering
    \includegraphics[width=\textwidth]{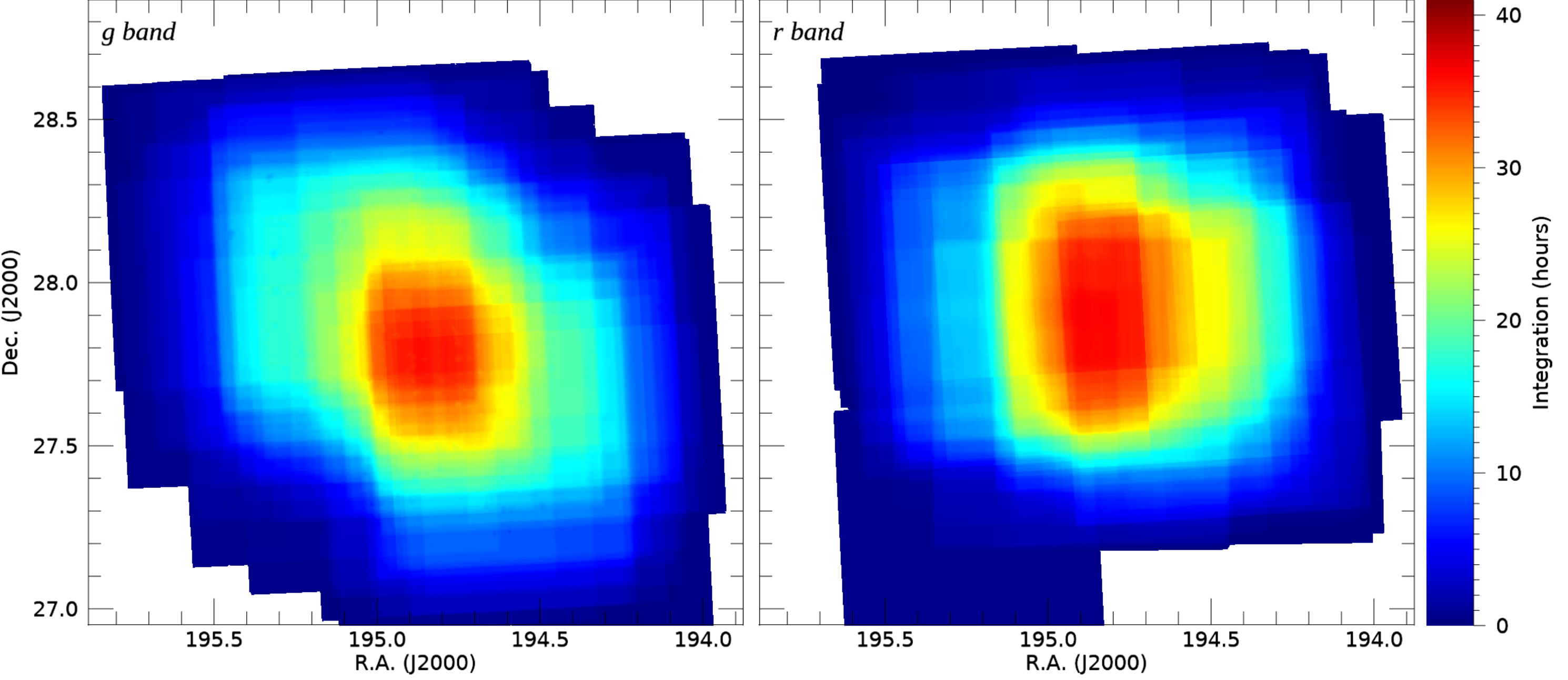}
    \includegraphics[width=\textwidth]{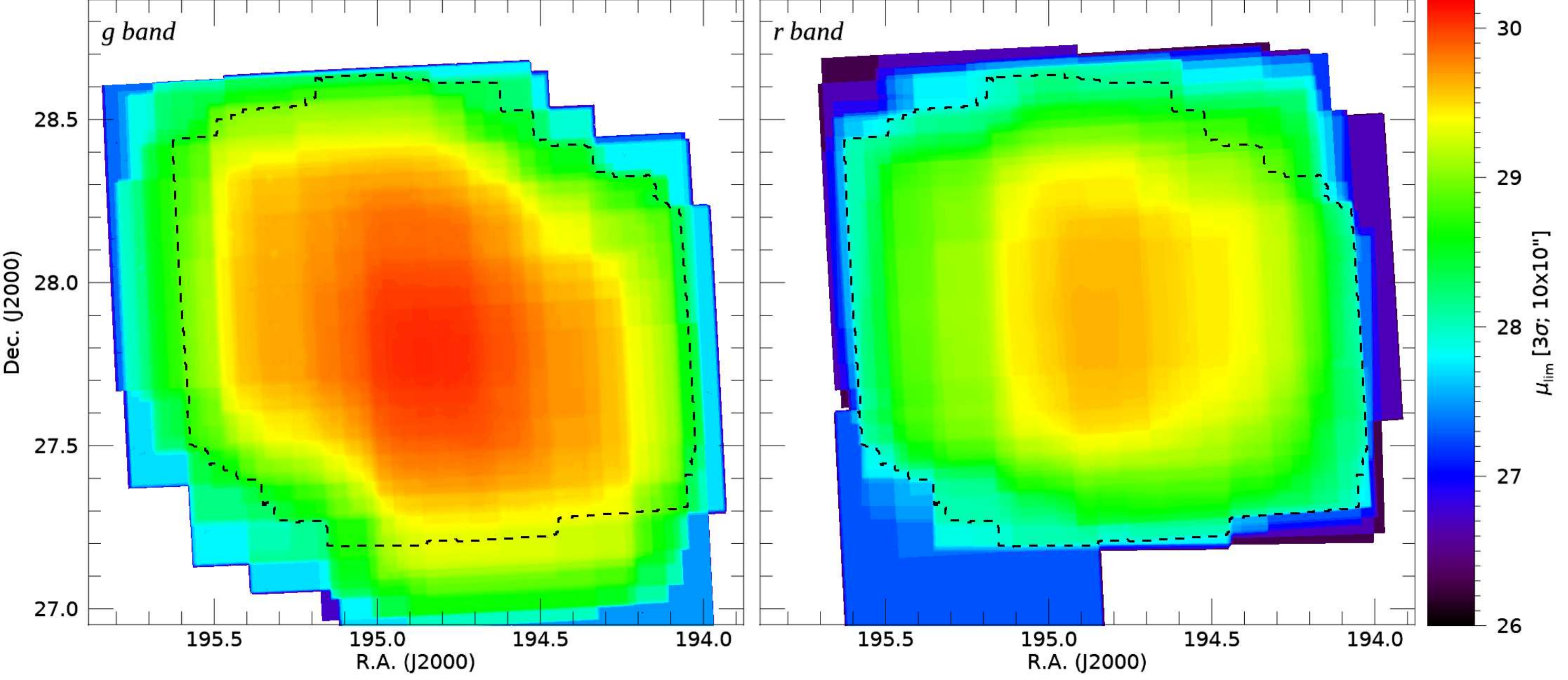}
    \caption{Upper panels: total integration-time maps across the HERON Coma footprint in the $g$ (left) and $r$ (right) bands. The color bars indicate the local integration time in hours. Lower panels: corresponding nominal limiting-surface-brightness maps, expressed as $3\sigma$ limits in $10\times10~\mathrm{arcsec}^{2}$ apertures. The dashed black contours delimit the common region with at least 1.5~h of integration in both bands adopted for the scientific analysis.}
    \label{fig:Mapas_exp}
\end{figure*}

For LSB imaging, the quality and reliability of a dataset cannot be adequately characterized by a single limiting surface-brightness value. It is important to distinguish between the nominal sensitivity set by random background fluctuations and the fidelity with which genuine extended astrophysical emission is preserved. The former, hereafter referred to as the nominal depth, describes the faintest surface brightness detectable against the statistical fluctuations of the background over a specified angular scale. In the ideal background-limited regime, the noise decreases approximately as $t_{\rm exp}^{-1/2}$ and therefore depends on the total exposure time, telescope collecting area, throughput detector sensitivity, and brightness of the night sky.

A separate limitation arises from systematic errors and astrophysical foregrounds. Imperfect flat-fielding, inaccurate modeling or subtraction of the sky background, instrumental reflections, and scattered light from the extended PSF wings of bright stars can generate artificial diffuse structures and can bias, attenuate, or even remove genuine LSB emission. In addition, diffuse Galactic cirrus constitutes an astrophysical foreground rather than a processing artifact. Its filamentary optical emission can overlap with or mimic tidal debris, stellar halos, and ICL, thereby limiting the reliable interpretation of the faintest detected structures.

These systematic effects cannot generally be reduced to a single universal quality metric and, unlike random noise, do not decrease as the square root of the exposure time. They can therefore become increasingly dominant as the random-noise floor is pushed to progressively fainter surface-brightness levels. By contrast, the nominal depth can be quantified reproducibly once both a significance threshold and an angular scale are specified. Throughout this work, we adopt the surface brightness corresponding to a $3\sigma$ background fluctuation in a $10\times10$~arcsec$^{2}$ aperture, following the standard metric adopted in previous low-surface-brightness studies \citep[e.g.,][]{Roman2020}. This quantity characterizes the random-noise sensitivity of the images, but should not by itself be interpreted as the practical limit for recovering extended emission. The optimal situation is one in which residual systematics and foreground contamination remain negligible relative to the random-noise floor on the spatial scales of interest, such that coherent astrophysical structures can be traced to surface-brightness levels consistent with the nominal depth. In the following, we quantify this nominal sensitivity across the HERON Coma footprint.

A direct measurement of the background noise at every position is complicated by the large angular extent of the Coma galaxies and the diffuse intracluster emission, which leave relatively few genuinely source-free regions in the deepest parts of the footprint. We therefore derived an empirical relation between local integration time and nominal limiting surface brightness.

The final coadds were first aggressively masked to retain only pixels dominated by the background. The remaining unmasked pixels were grouped according to their local integration time, as measured from the corresponding exposure map. For each integration-time interval and photometric band, we fitted a Gaussian function to the distribution of the individual pixel values and adopted the fitted width, $\sigma_{\rm pix}$, as the estimate of the per-pixel background noise. We found this estimator to be less sensitive than a directly calculated sigma-clipped standard deviation to residual non-Gaussian tails produced by imperfectly masked sources and other low-level contaminants. The measured per-pixel dispersion was subsequently scaled to the equivalent background fluctuation within a $10\times10~\mathrm{arcsec}^{2}$ aperture, following the procedure described by \citet{Roman2020}.

\begin{figure}
    \centering
    \includegraphics[width=\columnwidth]{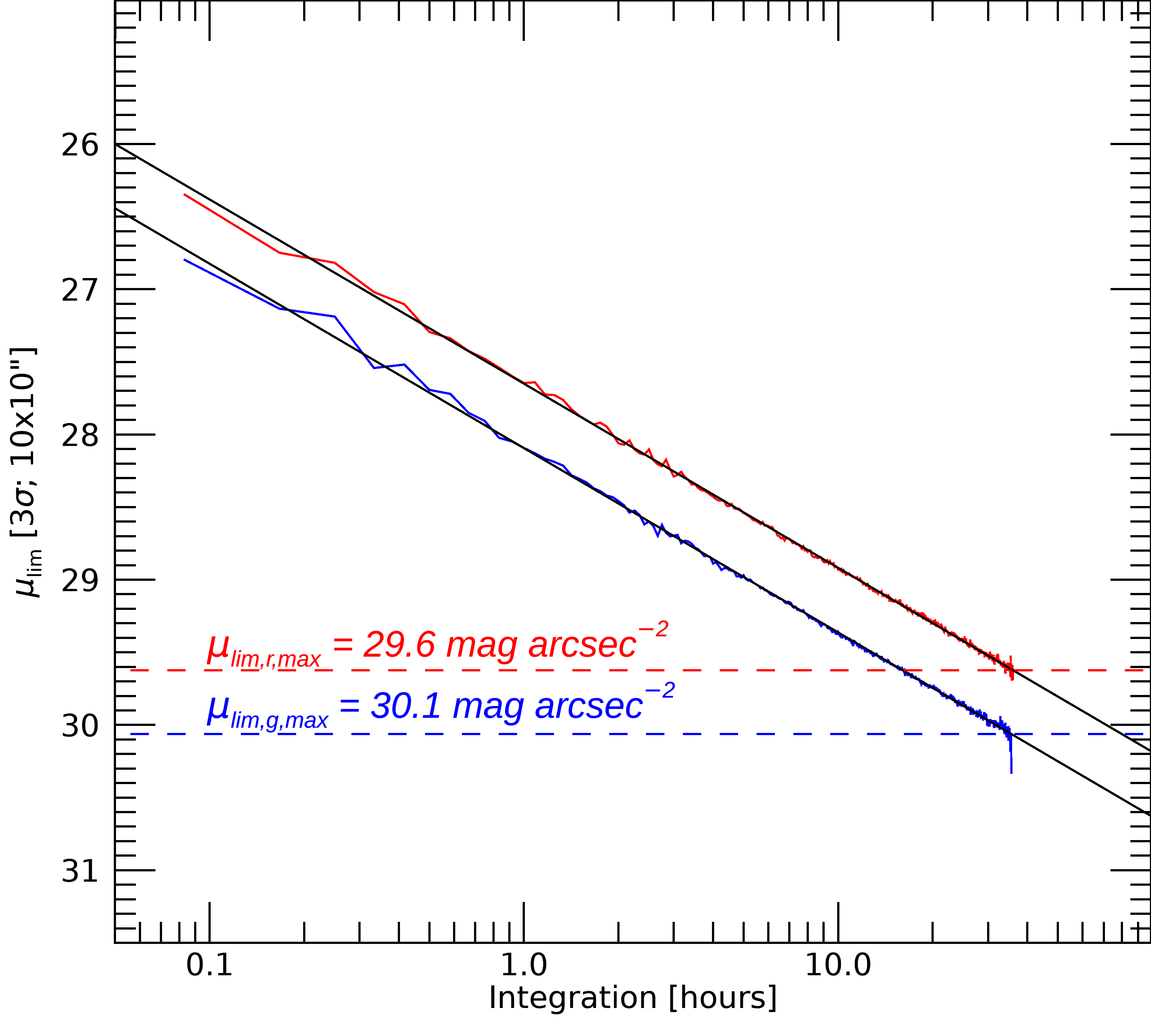}
    \caption{Nominal limiting surface brightness as a function of integration time. The blue and red curves show the measurements obtained for the $g$ and $r$ bands, respectively, while the solid lines represent the empirical relations given by Eqs.~\ref{eq:depth_g} and \ref{eq:depth_r}. The horizontal dashed lines indicate the maximum nominal depths inferred from the highest local integration time reached in each coadd.}
    \label{fig:Depth_plot}
\end{figure}

Figure~\ref{fig:Depth_plot} shows the resulting relation between integration time and nominal limiting surface brightness. The depth increases with integration time as expected for random noise decreasing approximately as $t_{\rm exp}^{-1/2}$. The empirical relations are

\begin{equation}
\mu_{\mathrm{lim},g}
=
-2.54\,\log
\left[
\frac{1}
{\sqrt{t_{\mathrm{exp}}(h)}}
\right]
+
28.09
\quad
\mathrm{mag\,arcsec^{-2}},
\label{eq:depth_g}
\end{equation}

\begin{equation}
\mu_{\mathrm{lim},r}
=
-2.54\,\log
\left[
\frac{1}
{\sqrt{t_{\mathrm{exp}}(h)}}
\right]
+
27.65
\quad
\mathrm{mag\,arcsec^{-2}}.
\label{eq:depth_r}
\end{equation}

The fitted coefficient of 2.54 is close to the value of 2.5 expected for ideal $t_{\rm exp}^{-1/2}$ noise scaling. All limiting surface brightnesses quoted here correspond to $3\sigma$ background fluctuations evaluated over a spatial scale of $10\times10~\mathrm{arcsec}^{2}$.

The limiting surface brightness cannot be measured directly at the highest integration times because the deepest regions are dominated by extended cluster emission and contain too few unmasked background pixels. The increasing scatter toward the deepest end of Fig.~\ref{fig:Depth_plot} reflects the decreasing number of available source-free pixels. We therefore inferred the maximum nominal depth by evaluating Eqs.~\ref{eq:depth_g} and \ref{eq:depth_r} at the highest local integration time reached in each mosaic.

Although the final coadds contain 458 $g$-band and 460 $r$-band exposures, the dithered pointings do not all overlap at a single position. The maximum local coverage is 427 exposures, corresponding to 35.6~h in the $g$ band, and 432 exposures, corresponding to 36.0~h in the $r$ band. The resulting maximum nominal depths are

\begin{equation}
\mu_{\mathrm{lim},g,\mathrm{max}}
=
30.1~\mathrm{mag\,arcsec^{-2}}
\quad
[3\sigma;\ 10\times10~\mathrm{arcsec}^{2}],
\label{eq:depth_g_max}
\end{equation}

\begin{equation}
\mu_{\mathrm{lim},r,\mathrm{max}}
=
29.6~\mathrm{mag\,arcsec^{-2}}
\quad
[3\sigma;\ 10\times10~\mathrm{arcsec}^{2}].
\label{eq:depth_r_max}
\end{equation}

These values are indicated by the horizontal dashed lines in Fig.~\ref{fig:Depth_plot}. Applying Eqs.~\ref{eq:depth_g} and \ref{eq:depth_r} to the integration-time maps provides the spatial distribution of the nominal limiting surface brightness across the footprint.

For the analyses presented in this work, we restrict the usable footprint to regions with a minimum integration time of 1.5~h in both $g$ and $r$. According to Eqs.~\ref{eq:depth_g} and \ref{eq:depth_r}, this threshold corresponds to nominal limiting surface brightnesses of approximately $28.3~\mathrm{mag\,arcsec^{-2}}$ in $g$ and $27.9~\mathrm{mag\,arcsec^{-2}}$ in $r$.

Figure~\ref{fig:Mapas_exp} shows the resulting spatial distributions of integration time and nominal limiting surface brightness. The central region has the deepest and most homogeneous coverage, while the sensitivity decreases progressively toward the boundaries of the mosaic. A substantial fraction of the central footprint reaches nominal depths of approximately $29.0~\mathrm{mag\,arcsec^{-2}}$ or fainter in both filters. The dashed black contours delimit the common analysis footprint adopted throughout this work.

\section{Method used to extract the NGC~4839 and NGC~4816 profiles}
\label{app:photometric_profiles_4839_4816}

The strongly boxy outer morphologies of NGC~4839 and NGC~4816 make simple elliptical apertures inadequate for characterizing their extended stellar envelopes. We therefore constructed customized isophotal apertures from empirical two-dimensional models designed to follow their observed light distributions.

We first applied aggressive masking to minimize contamination from surrounding sources. We combined the compact-source masks used to construct Figs.~\ref{fig:Coma_mag} and \ref{fig:Puente} with manually defined masks for larger objects. Extended circular masks were placed around potentially contaminating galaxies and other sources, reaching beyond their visually detected emission because the azimuthally averaged profiles probe substantially fainter levels. When measuring either target, we also masked the region occupied by the other galaxy to reduce contamination from its extended stellar envelope. For NGC~4839, the eastern region adjacent to the Coma ICL was additionally excluded. Nearly identical masks were applied in all bands and are shown as shaded regions in Fig.~\ref{fig:Perfiles}.

We then modeled the two-dimensional light distributions of both galaxies using \texttt{IMFIT} \citep{Erwin2015}. For each galaxy, the arithmetic-mean $\overline{g+r}$ image was fitted with three generalized S\'ersic components allowing boxy or disky distortions. These models were not intended to provide a physical structural decomposition, but to reproduce the observed morphology sufficiently well to define realistic isophotal apertures. The separate PSF correction applied before profile extraction is described below.

From these models, we constructed maps of generalized isophotal radius. Each pixel was assigned the radius of the model isophote passing through its position, allowing the radial bins to follow the observed morphology rather than a fixed family of ellipses.

Visual inspection of Fig.~\ref{fig:Puente} shows that the $\mu_{\overline{g+r}}\sim30~\mathrm{mag\,arcsec^{-2}}$ isophotes of NGC~4816 retain the boxy morphology reproduced by the \texttt{IMFIT} model. At a similar surface brightness, however, the outer isophotes of NGC~4839 become approximately elliptical. The adopted radius map for NGC~4839 therefore follows the boxy model geometry out to 400~kpc and transitions smoothly between 400 and 800~kpc to an outer elliptical geometry. The latter was fitted manually to the $\mu_{\overline{g+r}}\sim30~\mathrm{mag\,arcsec^{-2}}$ isophote in the deep $10\times10$ binned image shown in Fig.~\ref{fig:Puente}, yielding an axis ratio of $q=0.50$. A smooth radial interpolation between the angular shapes of the inner boxy isophotes and the outer ellipse preserves the well-constrained central morphology while avoiding irregular apertures in the outskirts.

Before extracting the profiles, we corrected the images for PSF-induced redistribution of light following an approach similar to that adopted in previous LSB studies \citep[e.g.,][]{TrujilloFliri2016, Gilhuly2022}. For each galaxy and band, we performed a separate \texttt{IMFIT} fit including the corresponding PSF model. From the best-fitting intrinsic model, we generated one image convolved with the PSF and another without convolution. Their difference estimates the net redistribution of light produced by the PSF and was subtracted from the observed $g$- and $r$-band images. The $\overline{g+r}$ image was corrected consistently using the single-band correction maps. After computing the PSF correction independently in $g$ and $r$, we constructed the corresponding $\overline{g+r}$ correction as their arithmetic mean and subtracted it from the original arithmetic-mean image. The upper panels and black surface-brightness profiles in Fig.~\ref{fig:Perfiles} are therefore based on PSF-deconvolved $\overline{g+r}$ images.

The photometric profiles were extracted in bins of generalized isophotal radius. We adopted logarithmic radial sampling, with an innermost bin containing the galaxy center and progressively spaced bins extending to the maximum radius. Pixels were required to be finite and unmasked simultaneously in the PSF-deconvolved $\overline{g+r}$, $g$, and $r$ images. This common-pixel selection ensures that the surface-brightness and color profiles sample exactly the same spatial regions.

Within each radial bin, sigma clipping was defined from the $\overline{g+r}$ image. The center of the flux distribution was initially estimated using the weighted median and its scatter using the median absolute deviation. Pixels deviating by more than $3\sigma$ were rejected iteratively, and the resulting selection was then applied unchanged to the $g$- and $r$-band images, preventing band-dependent clipping from biasing the color profile. The flux in each band was calculated as the weighted mean of the accepted pixels, with the exposure maps providing the relative weights. The effective number of pixels was estimated as

\[N_{\rm eff}=\frac{\left(\sum_i w_i\right)^2}{\sum_i w_i^2},\]

and the formal uncertainty on the mean flux was calculated from the weighted standard deviation divided by $\sqrt{N_{\rm eff}}$. Mean fluxes were converted into surface brightness using the adopted photometric zero point and pixel scale. The color profile was derived from the matched $g$- and $r$-band measurements, with its uncertainty obtained by propagating the corresponding errors in quadrature.

The profiles in Fig.~\ref{fig:Perfiles} should therefore be interpreted as PSF-deconvolved, matched-aperture measurements providing a first quantitative characterization of the extended light surrounding NGC~4839 and NGC~4816. Here, they illustrate the depth of the HERON data and their ability to trace the stellar envelopes of massive cluster galaxies to very large radii. A more detailed analysis of these profiles and their systematic uncertainties will be presented in a forthcoming paper dedicated to the NGC~4839--NGC~4816 complex.

\end{document}